\documentclass[aps,pra,onecolumn,notitlepage,superscriptaddress]{revtex4-1}

\usepackage{graphicx,color}% Include figure files
\usepackage{dcolumn}% Align table columns on decimal point
\usepackage{bm}% bold math
\usepackage{amsmath,amssymb,mathrsfs}
\usepackage{url}
\usepackage{hyperref}

\newcommand{\R}{\mathbb{R}}
\newcommand{\C}{\mathbb{C}}

\newcommand{\set}[1]{\mathsf{#1}}
\newcommand{\grp}[1]{\mathsf{#1}}

\def\d{{\rm d}}

\def\>{\rangle}
\def\<{\langle}
\def\kk{\>\!\>}
\def\bb{\<\!\<}
\newcommand{\st}[1]{\mathbf{#1}}
\newcommand{\bs}[1]{\boldsymbol{#1}}

\newcommand{\map}[1]{\mathcal{#1}}
\newcommand{\Tr}{\operatorname{Tr}}

\newcommand{\arccot}{\mathrm{arccot}\,}

\newcommand\myuparrow{\mathord{\uparrow}}
\newcommand\mydownarrow{\mathord{\downarrow}}
\newcommand\h{{\scriptstyle \frac 12}}

\newtheorem{theo}{Theorem}

\newtheorem{lemma}{Lemma}
\newtheorem{prop}{Proposition}

\newtheorem{prob}{Problem}

\def\Proof{{\bf Proof.~}}
\def\qed{$\blacksquare$ \newline}

\begin{document}
	\preprint{APS/123-QED}
    \title{Quantum-enhanced learning of rotations about an unknown direction}
    
    %\date{ \today}
        \author{Yin Mo} 
    \affiliation{Department of Computer Science, The University of Hong Kong, Pokfulam Road, Hong Kong}
    
     \author{Giulio Chiribella} 
      \affiliation{Department of Computer Science, The University of Hong Kong, Pokfulam Road, Hong Kong}
   \affiliation{Department of Computer Science, University of Oxford, Wolfson Building, Parks Road, Oxford, UK}

    \nopagebreak

	\begin{abstract} 
	%Learning machines   gather information from the  environment  and store it  in their internal memory. 
	  %   A fundamental  question is whether a machine equipped with  a quantum memory can learn more accurately  than all  machines equipped with a purely classical memory. Here we address this question for % answer the question in the affirmative, providing a concrete example of learning task exhibiting  a  gap between the optimal performance achievable with and without quantum memories. 
	     We design 
	      machines that learn  how to rotate a quantum bit about an initially unknown direction, encoded  in the state of a  spin-$j$ particle.
% In the learning process,  the machine sends out a probe, consisting of a spin-$j$ particle.   The direction is then imprinted into the probe,  and the corresponding information is transferred from the probe to the machine's internal memory. 
 %The direction is first imprinted into a probe, consisting of a spin-$j$ particle, and then is transferred from the probe to the machine's internal memory.  
   We show that a machine equipped with a quantum memory of $O(\log j)$ qubits   can outperform all  machines with  purely classical memory, even if the size of their memory is arbitrarily large.  The advantage is present for every finite $j$ and persists as long as the quantum memory is accessed   for no more than $O(j)$ times. We establish these results by deriving  the ultimate performance achievable with  purely classical memories, thus providing  a  benchmark that  can be used to experimentally demonstrate  the implementation of quantum-enhanced learning. 
	     
	       \end{abstract}
    \maketitle

	\section{Introduction}   
	
	Quantum machine learning \cite{biamonte2017quantum, dunjko2018machine}  explores the interface between machine learning and quantum information science.   On the one hand, quantum algorithms    have been shown to offer  speedups to a variety of  classical machine learning tasks \cite{aimeur2006machine, harrow2009quantum, rebentrost2014quantum, ronnow2014defining, wiebe2014quantum, dunjko2016quantum, amin2018quantum}.  On the other hand, ideas from machine learning stimulated   the formulation of  new quantum  tasks, such as  quantum state classification \cite{sasaki2001quantum, sasaki2002quantum, guctua2010quantum, sentis2012quantum},  quantum learning of gates \cite{bisio2010optimal,marvian2016universal,sedlak2018optimal}  and measurements \cite{bisio2011quantum, sentis2012quantum}.

 An important component of any learning machine is its internal memory, wherein   information gathered  from the environment is stored. 
 % In general, the memory can consists of both classical and quantum bits. 
 %which in general can consist of both classical and quantum bits. 
  % wherein   information gathered  from the environment is stored. 
 % until needed for the execution of a prescribed task.  
 A machine equipped with purely classical memory can only gather information through measurements, and can only perform conditional operations  controlled by  classical data.     
 %We call this approach  {\em measure-and-operate}. 
      In contrast, a   machine equipped with a quantum memory can gather information by  interacting coherently with its environment, and can perform operations that are controlled by quantum data. 
    %  A quantum learning machine can be viewed as a special type of  programmable quantum device \cite{nielsen1997programmable,vidal2002storing,hillery2002probabilistic,vidal2002storing,brazier2005probabilistic,ishizaka2008asymptotic}, which synthesises its program through interaction  with the surrounding environment.    
A fundamental question is whether the additional freedom offered by the quantum memory  can enhance the learning performance.

A task where  quantum memories are known to enhance the performance  is quantum cloning \cite{scarani2005quantum}, which can be rephrased as the task of  learning  how to prepare copies of an unknown quantum state by gathering sample copies of it.   In this task, a machine with a quantum memory can achieve   strictly higher accuracy than all machines with a purely classical memory \cite{buvzek1996quantum,gisin1997optimal,werner1998optimal}.
% as long as the number of output copies is finite \cite{bae2006asymptotic,chiribella2006quantum,chiribella2010quantum,chiribella2014optimal,gendra2014probabilistic}. 
      
       A strikingly different situation occurs in the task of learning how to perform unitary gates.  In this case, quantum memories can enhance the performance of probabilistic learning machines \cite{nielsen1997programmable,vidal2002storing,hillery2002probabilistic,vidal2002storing,brazier2005probabilistic,ishizaka2008asymptotic,bartlett2009quantum,sedlak2018optimal}, but the enhancements  observed so far disappear if the machines are required to approximate the desired gate with unit probability \cite{bisio2010optimal}.   The reason for such behaviour is that the learning machines considered so far  were designed to perform   {\em groups} of unitary gates, such as the group $\grp {SO} (3)$ of all qubit rotations, or the group $\grp U(1)$ of qubit rotations about a fixed axis.    In these highly symmetric scenarios,      a general theorem    by Bisio {\em et al } \cite{bisio2010optimal} implies that every quantum machine operating with unit probability can be replaced by  a machine that achieves the same learning accuracy with a purely classical memory. 
          Given the generality of this result,  one may be tempted  to conjecture that quantum memories are of no use for  deterministically learning  unitary gates.  Such  conjecture would be    consistent with Nielsen and Chuang's {\em no-programming theorem} \cite{nielsen1997programmable}, which implies that whenever a set of unitary gates  can be perfectly encoded into  a set of quantum states, the states in the set must be  orthogonal, and therefore storable in a  purely classical memory.

In contrast with the above  observations, here we show that quantum memories can generally enhance the performance of deterministic  machines attempting to learn  unitary gates. To make this point, we provide a concrete example where  the optimal deterministic  learning strategies, with and without quantum memory, can be determined explicitly. 
%Comparing the performances of the optimal learning strategies with and without quantum memory, we establish that quantum memory 
Specifically, we consider machines that learn  how to rotate a quantum particle by a desired angle $\theta$ about an initially unknown direction $\st n  =  (n_x,n_y, n_z)$,  imprinted in the state of a  spin-$j$ particle.   
%Since the rotation angle is fixed, the target operations do not form a group, and the advantage of the quantum memory in this scenario is not in contradiction  with Bisio {\em et al}'s no go theorem  \cite{bisio2010optimal}.
 % Accordingly, we find that deterministic machines equipped with a quantum memory offer advantages for all rotation angles  $\theta$ and for all spins  $j>1$.   For spins $j=1$  and $j=1/2$, quantum memories still offer an advantage for rotation angles away from $\pi$,  but the advantage disappears as  $\theta$ approaches $\pi$. 
   We  consider two different ways of imprinting the direction in a spin-$j$ particle, corresponding to the following scenarios: 
%  {\em (1)} resetting the probe's state to a  state pointing in direction $\st n$,   and {\em (2)} subjecting the probe to a unitary gate that rotates the direction of the $z$-axis into the direction $\st n$.    
%These two ways of imprinting the direction  arise naturally in   the following scenarios:  
   \begin{itemize}
 \item {\em Scenario 1: spin relaxation. }   
  A static   magnetic field $\st B   =  (B_x,B_y,B_z)$, pointing in an unknown direction   $\st n  =  \st B/\|\st B\|$, acts in a certain region of space.   
%  The  task of the learning  machine is rotate a target particle, sitting outside  region $R$, about the field's     To this purpose, the
%A machine sends out 
A spin-$j$   probe  enters  the region and    undergoes a thermalisation process with respect   to the magnetic dipole Hamiltonian $  H=  -\mu  \, \st B  \cdot  \st J$, where  $\st J  =  (J_x,J_y,J_z)$ are  the spin operators,  $\st B  \cdot  \st J  :  = B_x J_x   +  B_y  J_y  +  B_z  J_z$, and $\mu>0$ is a suitable constant. For simplicity, we   assume that the temperature is low enough that the thermal state  is approximately the ground state of the Hamiltonian, namely the eigenstate of the operator $\st n\cdot  \st J: =  n_x  J_x  +  n_y J_y  +  n_z  J_z$  with  maximum eigenvalue $j$, hereafter denoted as $|j,j\>_{\st n}$.    
  An   extension to thermal states at finite temperature will be discussed in  Section \ref{sec:robustness}.

  % This state is known as a  spin coherent state \cite{arecchi1972atomic, perelomov2012generalized} and is generally regarded as a canonical example of  quantum state indicating a  direction  \cite{bartlett2007reference}.    
   
   \item {\em Scenario 2: action of  an unknown unitary gate.}  A  black box    implements an unknown rotation $g  \in  \grp{SO} (3)$, which transforms the $z$-axis into the direction $\st n $.  %The task of the learning machine is to implement a rotation by an angle $\theta$ about the direction $\st n$.   To this purpose, 
   A machine prepares a spin-$j$ probe in an initial state  $|\phi_\theta\>$  (possibly depending on the desired rotation angle), and sends the probe as input to the black box.  %Before sending the probe, the machine entangles the probe with an auxiliary system that is unaffected by the rotation.  
   %After the action of the black box, the joint state of the probe and the auxiliary system will be $|\phi_{\theta,g}\>  =  \left( U_g^{(j)}\otimes I_{\rm A}\right) |\phi_\theta\>$, where $|\phi_\theta\>$ is the initial state (possibly depending on the desired rotation angle $\theta$), $I_{\rm A}$ is the identity on the auxiliary system, and $U_g^{(j)}$ is the unitary matrix representing the rotation $g$ on a spin-$j$ particle. 
   After the action of the black box, the state of the probe is $|\phi_{\theta,g}\>  =  U_g^{(j)}   |\phi_\theta\>$, where  $U_g^{(j)}$ is the unitary matrix representing the action of the rotation $g$. 
                  \end{itemize} 
        Scenario 2  is also relevant to the study of quantum reference frames  \cite{bartlett2007reference}.  Our learning task can be translated into a distributed quantum protocol  involving two distant parties, Alice and Bob, who do not share a reference frame for spatial directions. The goal of the protocol is to allow Bob to rotate a target particle by a desired angle $\theta$ about the direction of Alice's $z$-axis, encoded in the state of a spin-$j$ particle prepared by Alice and sent to Bob as a token of her reference frame.  In this setting, the unknown rotation $g$ describes the mismatch between Alice's and Bob's Cartesian axes, and the optimal learning strategy provides the optimal protocol for encoding  the direction of Alice's $z$-axis in a spin-$j$ particle and for rotating Bob's target particle accordingly. 
              
                A key difference between Scenarios 1 and 2 is that   the initial state of the probe is irrelevant in Scenario 1 (every initial state is  reset to the %spin coherent 
                state $|j,j\>_{\st n}$), while  it can be optimised in Scenario 2.   
                %In principle,
                More generally,  the optimal probe state in Scenario 2 could  be an entangled state involving, in addition to the the spin-$j$ particle,  an auxiliary system stored in the internal memory of the machine.  Nevertheless, we will show that such auxiliary system does not increase the accuracy in the execution of the desired rotation, and therefore it can be omitted without loss of generality.  
                %, and that the optimal accuracy in the execution of the desired rotation can be achieved without using any auxiliary system.  

 %  Scenarios 1 and 2  share a common mathematical structure.  In both cases, the information about the target operation is imprinted in a state of the form $ \left( U_g^{(j)}\otimes I_{\rm A}\right) |\phi_\theta\>$, for some initial state $|\phi_\theta\>$ and some unknown rotation $g  \in  \grp{SO} (3)$.   Scenario 1 corresponds to the special case where the auxiliary system is trivial ({\em i.e.} one-dimensional) and the initial state $|\phi_\theta\>$ is  $|j,j\>$,  the eigenstate  of $J_z$ with maximum eigenvalue $j$. 
%   Indeed, a generic spin coherent state  $|j,j\>_{\st n}$   can be represented as $|j,j\>_{\st n}  =  U_{g(\st n)}^{(j)}  |j,j\>$, where $g(\st n)$ is a rotation that aligns the $z$-axis with the direction $\st n$  \cite{perelomov2012generalized}.     Taking advantage of this fact, we show that the optimal  strategy for Scenario 1 coincides with the optimal strategy for Scenario 2 under the constraint  $|\phi_\theta\>  =  |j,j\>$. 
   
In this paper we establish the optimal learning strategies for both Scenarios 1 and 2, focussing on the case where the target particle is a qubit. 
   %We will start from Scenario 2, as it corresponds to a more general optimisation problem.   %Our first result is that no reference system is necessary and that the initial state $|\phi\>$ can be chosen without loss of generality to be $|j,m\>$,   the eigenstate  with eigenvalue $m$ of the $z$ component of the angular momentum operator. 
  A summary of the key result is as follows. For $j>1$, we find that the optimal strategies for Scenarios 1 and 2 coincide. 
%For every desired angle $\theta$, the optimal probe state in Scenario 2 is $|\phi_\theta\> = |j,j\>$, the eigenstate of $J_z$ with maximum eigenvalue $j$. As a consequence, the output state  after the action of the black box is $|\phi_{\theta,g}\>  =  |j,j\>_{\st n}$, where $\st n$ is the direction of the rotated $z$-axis.  The optimal     strategy consists in
% show that  the optimal strategy for Scenario 2 consists in  
In both cases, the optimal learning strategy consists in 
\begin{enumerate}
\item preparing the probe in the  initial state $|\phi_\theta\>  =  |j,j\>$, the eigenstate of $J_z$ with maximum eigenvalue $j$
\item imprinting the direction in the probe, and storing the resulting state in a quantum memory of $\lceil \log (2j+1)\rceil $ qubits
\item retrieving the probe's state from the memory and  letting it interact with the target through the isotropic Heisenberg interaction  $H  \propto     \sigma_x   J_x  +  \sigma_y J_y  +  \sigma_z J_z$, where $(\sigma_i)_{i=x,j,z}$ are the Pauli matrices for the target qubit. 
% and $(J_i)_{i=x,y,z}$ are the  components of the spin operator of the memory. 
\end{enumerate}   

Notably, the structure of the optimal learning machine is independent of the desired rotation angle $\theta$:   a single probe state and a single interaction Hamiltonian work optimally for all possible  angles. 
 The  rotation angle    only affects the interaction time between the probe and the target. 
 
For every  $j>1$, we prove that the optimal machine with quantum memory  outperforms every machine with purely classical memory.  We  determine  the optimal fidelity over all machines with purely classical memory, providing a  benchmark that can be used  to  demonstrate the  advantage of  quantum memories in realistic experiments.   For example,  we show that  the optimal classical strategy   for $j=3/2$ and $\theta=  \pi$  has fidelity   $64\%$, while  the optimal quantum strategy has  fidelity of  $71\%$.               As a consequence,   every  experimental fidelity above $64\%$ guarantees the  demonstration of  quantum-enhanced learning.  
 In general, we  show that  a non-zero quantum advantage is present for every rotation angle  $\theta \not = 0$ and for every $j>1$. We also prove that the advantage persists even if the memory is accessed multiple times,  as long as the number of accesses to the memory is $O(j)$.   In Scenario 1, we find that the quantum advantage persists  at non-zero temperature $T$, %when the probe state is reset to a thermal state. 
 as long as the magnetic energy $\mu  \|\st B\|$ is large compared to the thermal fluctuation $k_{\rm B}  T$, $k_{\rm B}$ being the Boltzmann constant. 
 %,  the quantum memory offers a provable advantage. 

%The amount of the advantage, however, decreases in the large $j$ limit, in which both  MO  strategies  and general quantum strategies can implement the target gate with error vanishing as $O(1/j)$.  
	 
%Nevertheless, the error of the optimal quantum strategy vanishes twice as fast as the error of the optimal classical strategy. 

For  $j=1$, we find out a striking difference between  Scenarios 1 and 2. 
In  Scenario 1, the quantum memory offers an advantage  for all possible rotation angles. In  Scenario 2, the advantage disappears when the rotation angle approaches $\pi$. In that regime,  the optimal strategy  consists in  
\begin{enumerate}
\item preparing the probe in the  initial state $|\phi_\theta\>  =  |1,0\>$, the eigenstate of $J_z$ with eigenvalue $m=0$, 
\item sending the probe to the unknown gate $U_g^{(j)}$, and measuring  the resulting state $U_g^{(j)} |1,0\>$ on  the basis $\{|1,0\>_i \}_{  i \in \{x,y,z\}}$, where $|1,0\>_i$ is the eigenstate of $J_i$ with eigenvalue $m=0$, 
\item  conditionally on  outcome  $i$, performing a spin flip around the $i$-axis. 
  \end{enumerate}   
 
 For $j=1/2$, the optimal strategies for Scenarios 1 and 2 coincide, and the availability of a quantum memory offers  advantages  for all  rotation angles except $\theta = 0$ and $\theta=  \pi$.   

	The paper is structured as follows.
	In Section \ref{sec:intro} we introduce the problem of learning a rotation about an unknown direction, considering two alternative ways of imprinting the direction into the state of a spin-$j$ probe.  
	We derive the optimal quantum strategy in Section \ref{sec:accuracysize}, and the corresponding quantum benchmark in Section \ref{sec:benchmark}.
    In Section \ref{sec:longevity}, we show that the advantage persists even if the memory state is accessed multiple times,
    and in Section \ref{sec:robustness}, we show that the optimal learning strategy for Scenario 1 is robust to thermal noise.
    In Section \ref{sec:larger}, we extend our results from qubits to systems of arbitrary dimensions.   The conclusions are drawn in Section \ref{sec:conclusions}.

	\section{Learning how to rotate about an unknown axis} \label{sec:intro} 
	In this section we introduce the task of learning how to rotate a quantum particle about an initially unknown axis. We  consider two scenarios, in which the unknown axis is imprinted  in a quantum probe via two physically different processes:  (1) spin relaxation, 
	and (2) action of an unknown rotation gate.
	We formalise the optimisation problems corresponding to these scenarios and establish a relation between the corresponding solutions. 
	% In the first scenario,  the goal is to simulate the precession of a particle in  an unknown  field. In the swhich can be probed by a spin-$j$ particle. In the second scenario, the goal is to  Then, we show that the problem of learning a rotation can be viewed as an instance of a  more general learning task, where a machine has to learn how to perform a unitary gate $V_x$ by observing the action of another unitary gate $U_x$,  $x$ being  an  unknown parameter. 
	
	\subsection{Scenario 1: learning from a relaxation process}\label{subsect:precession}
	Suppose that a   static  magnetic field $\st B   =  (B_x,B_y,B_z)$  is turned on for a limited amount of time in a bounded region of space. While the field is turned on, a spin-$j$ particle is  placed in  the region and   undergoes a relaxation process, whereby its spin becomes aligned with the field's direction. % similarly to the needle of a compass.
	% in the same way as the needle of a compass aligns itself with the direction of the Earth's magnetic field. 
  %which after a transient phase  ends up pointing in the direction of the external field.  
   After the alignment has taken place, the state of the particle is  stored in the internal memory of a quantum machine, which will later use it to rotate a target particle by a desired angle $\theta$  about the direction $\st n  =  \st B/\|  \st B\|$.  
	%to simulate  the dynamics that a test particle would have experienced   {\em if it were immersed in the original field}.   
	
	%Or even if the field is still on, the memory could be used to reproduce the dynamics at a different location, where the field is not present. 
	We denote the spin-$j$ particle as  ${\rm P}_{j}$, and let  $J_x, J_y$ and $J_z$ be its spin operators, satisfying the commutation relations $[  J_x, J_y]   =  i J_z$,   $[  J_y, J_z]   =  i J_x$,    and $[  J_z, J_x]   =  i J_y$. All throughout the paper  the standard notation $|j,m\>$   (respectively, $|j,m\>_{\st n}$) for the eigenstate of the operator $J_z$  (respectively, $\st n \cdot \st J$) with eigenvalue $m $. 
	
	The  alignment of the magnet with the external magnetic field   can be described by a thermalisation process, whereby the initial state of the magnet converges to the thermal state of  the  magnetic   Hamiltonian  $  H  =-  \mu  \,   \st B \cdot  \st J  =  -\mu      (B_x J_x  +  B_y J_y +  B_z  J_z)$, where $\mu>0$ is a suitable constant.     For simplicity,  we will  assume that %the interaction with the thermal bath lasts  long enough that the final state is well approximated by the thermal state $\rho_{\rm th}  \propto  \exp [  -  H/(k_{\rm B} T)]$, and that 
	the temperature of the bath is low enough  that the thermal state is approximately the ground state of $H$.  %An extension to thermal states at finite temperature will be discussed in Section \ref{sec:robustness}. 
	Explicitly, the ground state is the spin coherent state $|j,j\>_{\st n}$.

	Overall, the alignment process can be modelled as a quantum channel (completely positive trace-preserving map) $\map T_{\st n}$ that resets every state of the probe to the  state $|j,j\>_{\st n}$.   In Section \ref{sec:robustness} we will extend our discussion to the finite-temperature scenario, where the channel $\map T_{\st n}$ resets the probe state to the thermal state of the magnetic dipole Hamiltonian.

	The goal  of the quantum machine is to rotate a target particle  $\rm S$  by a given angle $\theta$ about the direction  $\st n$.  We will mostly focus on the case where the target is a spin-$1/2$ particle,  regarded as a qubit. 
	In this case, we denote the target rotation by  $V_{\theta, \bf n}  :=   \cos \frac \theta 2   \, I  -  i  \sin \frac \theta 2 \,  \st n\cdot \bs \sigma $, where $\bs \sigma =  (\sigma_x,\sigma_y,\sigma_z)$ are the three Pauli matrices, and $\st n\cdot \bs \sigma : =  n_x \sigma_x  +  n_y\sigma_y +  n_z\sigma_z$. 
   	To  learn how to implement the target rotation, the machine will  transfer information from the magnet to its internal memory $\rm M$. Mathematically, this operation is described by a quantum channel  (completely positive trace-preserving map) $\map E_\theta$  transforming states of ${\rm P}_j$  into states of $\rm M$.   To be completely general, we allow the channel to depend on the desired  angle $\theta$. 
	    If the memory is classical,  the channel $\map E_\theta$ represents a measurement on the magnet, followed by the storage of the outcome in the  memory.  If the memory is quantum,  the channel $\map E_\theta$ 
	can be any process  transforming states of the magnet into states of the memory.   
%      If the memory is classical, its states can be represented as density matrices that are diagonal in a given basis, and the channel $\map E_\theta$  must be  of the form 
%	\begin{align}
%		\map E_\theta  (\cdot)    = \sum_{y\in\set Y}  \,       \Tr[ P_{\theta, y} \,   \cdot ]\, |y\>\<y|  \, ,	
%	\end{align}
%	 where  $\{  |y\>\}_{y\in\set Y}$ is the given basis for the Hilbert space of  the memory,   and  $(P_{\theta, y})_{y\in\set Y}$ is a Positive Operator-Valued Measure (POVM), that is, a set of operators satisfying the conditions   $P_{\theta, y} \ge 0 \, , \forall y\in\set Y$ and $\sum_{y\in \set Y}   P_{\theta, y}  =  I_{ {\rm P}_j}$, where $I_{{\rm P}_j }$ is the identity operator on the Hilbert space of the particle. 
  	 	
	When asked to perform the target rotation, the machine will retrieve information from  its internal memory,   and will use such information to control the evolution of the target system, hereafter denoted by $\rm S$. 
	If  the memory is classical, the control amounts to a conditional operation on the target depending on the classical data stored in the memory. If the memory is quantum, the control can be any general interaction between the memory and the target system.     In both cases, the control operation can be  described by a quantum channel $\map R_\theta$ transforming joint states of the composite system $\rm M\otimes \rm S$ into states of $\rm S$. 
	
\begin{figure}[h!]
		\centering
		\includegraphics[width=0.82\textwidth]{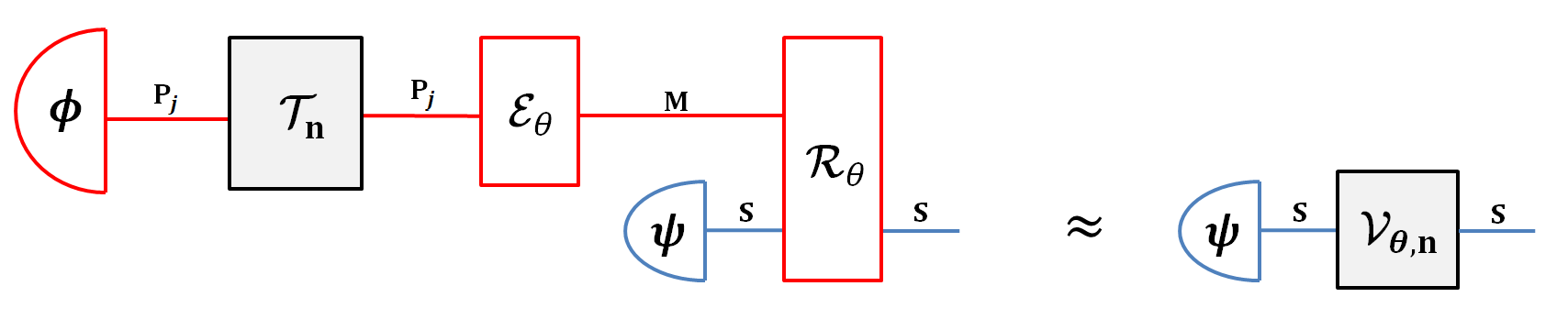}
		\caption{
			\textbf{Learning  from a relaxation process.}   {}	
			A spin-$j$ particle, initially in the state $|\phi\>$,  undergoes a relaxation process $\map T_{\bf n}$, which aligns its spin with the direction $\bf n$ of an external magnetic field.
			Information about the direction is then transferred from the spin-$j$ particle into the machine's internal memory $\rm M$.  The task of the machine is to rotate  a target qubit $\rm S$ by an angle $\theta$ about the direction $\st n$. To this purpose, the machine will perform a joint  operation $\map R_{\theta}$ on its internal memory and on the target, designed to approximate  the desired unitary gate $V_{\theta, {\bf n}}$. }  
		\label{learn_Thermal}
	\end{figure}

	Overall, the structure of the learning process is depicted in Figure \ref{learn_Thermal}. 
	If the initial state of the target is $|\psi\>$, then the final state is
	\begin{align}
		\nonumber \rho_{\theta,\st n}    &=      \map R_\theta  \Big (    \map E_\theta \left(  |j,j\>_{\st n}  \<j,j|_{\st n}  \right) \otimes |\psi\>\<\psi|   \Big) \\
		&  =   \map C_\theta  \big (     |j,j\>_{\st n}  \<j,j|_{\st n}  \otimes |\psi\>\<\psi|   \big) \, , 
	\end{align}
	where $\map C_\theta    :=  \map R_\theta  \circ (\map E_\theta \otimes \map I_{\rm S})$   is the effective quantum channel transforming joint states of the probe and the target into states of the target alone. 

	To evaluate the accuracy of the learning process, we  compare the output state $\rho_{\theta,\st n}$   with the desired  output  $V_{\theta, \bf n} |\psi\>\<\psi| V_{\theta, \bf n}^\dag $. As a figure of merit, we use 
	the {\em average input-output fidelity}  \cite{gilchrist2005distance} 
    \begin{align}\label{Fave1}
        F_1(j,\theta) :=  \int {\rm d} {\bf n}  \,  \int \d\psi 
        \, \<\psi|V_{\theta,\bf n}^{\dag}  \,  \Big[  \mathcal{C}_{\theta} \big(|j,j\>_{\st n}\<j,j|_{\st n}  \otimes  |\psi\>\<\psi|   \big)\Big]  \,  V_{\theta,\bf n}|\psi\> \, , 
    \end{align}
    where ${\rm d} \bf n$ is the rotationally-invariant probability distribution on the unit sphere,  
    $|\psi\>$ is  the initial state of the target qubit, and  $\d \psi$ is the unitarily invariant probability distribution on the pure states.    The associated  optimisation problem  is: 
	\begin{prob}\label{prob:channelonly}
		Find the  quantum channel $\map C_\theta$  that maximises the fidelity  $F_1(j,\theta)$ in Equation (\ref{Fave1}).
	\end{prob}  
	The optimisation can be performed with different constraints on the channel $\map C_\theta$, corresponding to different assumptions on the machine's internal memory.  In this paper, we will consider two cases: 
	\begin{enumerate}
		\item The machine is equipped with a quantum memory of     $\log  \lceil 2j+1\rceil$ qubits. In this case, the  channel $\map C_\theta$ is an arbitrary completely positive trace-preserving map. 
		\item The machine is equipped with a  classical memory of arbitrary size.   In this case, the channel  $\map C_\theta$ must be decomposable into a measurement on the probe followed by a conditional operation on the target. 
	\end{enumerate} 
	We will carry out both optimisations and compare the maximum fidelity achievable with a quantum memory with the maximum fidelity achievable with classical memories of arbitrary size.

   	\subsection{Scenario 2:  learning from a rotation gate}

Consider the following general problem. A  quantum machine has access to one use of a black box implementing some unknown unitary gate $U_x$, randomly drawn from some set $(U_x)_{x\in\set X}$.  By interacting with the black box, the machine has to learn how to perform another  unitary gate $V_x$, acting on a target system $\rm S$.   Typically, the  gate learning problems considered so far correspond to the case $V_x  =  U_x$  (the machine attempts to emulate  the gate $U_x$  \cite{bisio2010optimal,marvian2016universal,sedlak2018optimal}), or to the case $V_x  =  U_x^\dag$ (the machine attempts to invert the gate $U_x$ \cite{bartlett2009quantum,bisio2010optimal}).   In general, the relation between $U_x$ and $V_x$ can be arbitrary. 
 
To learn the target gate, the machine sends a probe $\rm P$ to the black box.  In general, the probe can be   entangled with an auxiliary system $\rm A$, stored in the machine's internal memory.    If the initial state of the composite system $\rm{ P} \otimes {\rm   A}$ is $|\phi\>$, then the state after the action of the black box is 
	\begin{align}\label{learning}
		|\phi_x\>  :=  (  U_x  \otimes  I_{\rm A})  \,  |\phi\>  \, ,
	\end{align}
	where $I_{\rm A}$ denotes the identity operator on the auxiliary system.

After the black box has acted, the probe returns to the machine, which transfers information from the state $|\phi_x\>$  to its internal memory $\rm M$.  The transfer of information is described by a quantum channel  $\map E$ with input system $\rm P \otimes \rm A$ and output system $\rm M$. Overall, the imprinting of the parameter $x $  in the machine's memory is  called the {\em training phase}.  Accordingly, we call $U_x$ the {\em training gate}.

\iffalse  In general, the training phase could consist of a sequence of operations in which the machine gains information about the parameter $x$ through multiple rounds of interaction with the gate $U_x$  \cite{bisio2010optimal}. Here, however, we  limit our attention to a single round, possibly with a  large multipartite probe. This  formulation already includes several relevant scenarios, such as learning how to perform a completely unknown gate by observing its action  \cite{nielsen1997programmable,vidal2002storing,hillery2002probabilistic,vidal2002storing,brazier2005probabilistic,ishizaka2008asymptotic,bisio2010optimal,sedlak2018optimal}, or learning how to invert a completely unknown gate \cite{bartlett2009quantum,bisio2010optimal}. 
  The first set of examples corresponds to the setting $U_x  =  V_x^{\otimes N}$, while the second set corresponds to the setting 
    $U_x  =   V_x^{\dag \, \otimes N}$.     In all these examples, the size of the training data is quantified by the total dimension of the probe, which  consists of $N$ identical subsystems. 
\fi 
	
	After the training phase has been concluded, the machine will be asked to perform the gate $V_x$ on the target system.  
We call this phase  the {\em execution phase}.    The  machine will access its internal memory and use it to control the evolution of the target system.  The control mechanism  is described by a quantum channel $\map R$ with input system  $\rm M \otimes\rm S$  and output system $\rm S$. 
	
\begin{figure}[h!]
	\centering
	\includegraphics[width=0.85\textwidth]{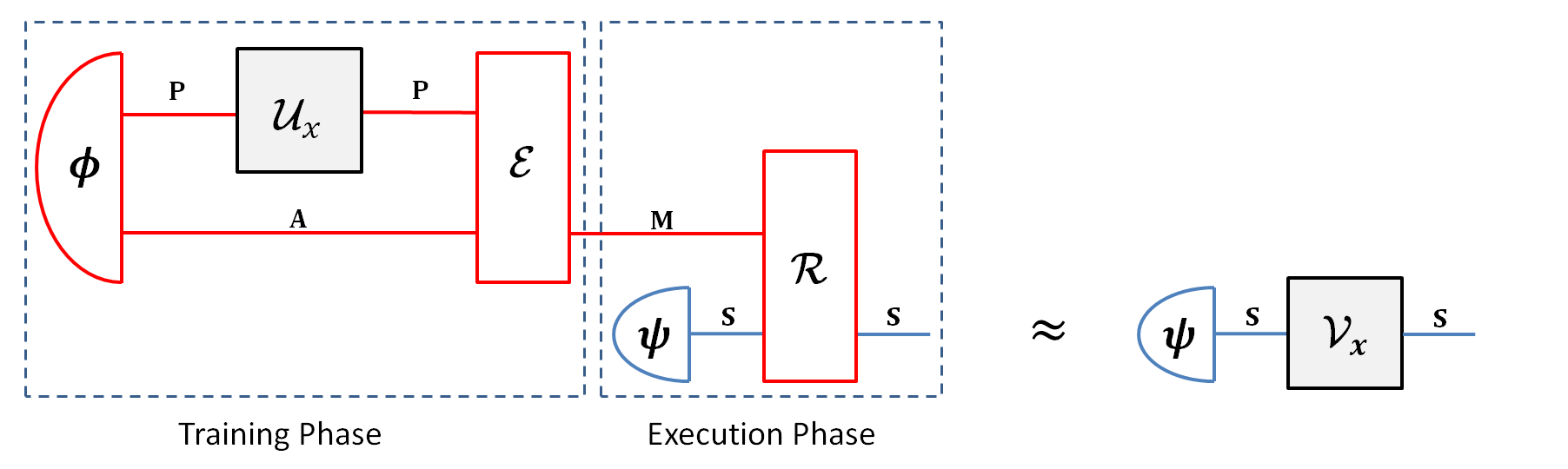}
	\caption{		\textbf{Learning from a unitary gate: the $U_x$-to-$V_x$  learning problem.} 
		A machine learns to perform a target gate $V_x$ by probing  a training gate $U_x$.
		In training phase, a probe is prepared together with an auxiliary system in a joint state $|\phi\>$.  The probe undergoes the gate $U_x$, and the  joint state of the probe and auxiliary system becomes $|\phi_x\>  :  =  (U_x\otimes I_{\rm A})  |\phi\>$.   Information from the state  $|\phi_x\>$ is then  transferred to the  machine's internal memory through an encoding channel $\map E$. In general, the memory can be quantum, classical, or a hybrid quantum-classical system.
		% If the memory is purely classical, encoding map $\map E$ will amount to a measurement, whose outcome is  stored in the classical memory.    
		In execution phase, machine performs a joint operation $\map R$ accessing its internal memory and use it to try to perform $V_x$ on the target system.}
	\label{learn_x}
\end{figure}  

% The $U_x$-to-$V_x$ learning problem is to find the optimal input state $|\phi\>$ and  the optimal channels $\map E$  and $\map R$ in order to maximize the accuracy in the execution of the gate $V_x$.  If the machine is equipped with a sufficiently large quantum memory, the channel $\map E$ can be taken to be a faithful encoding of the state $|\phi_x\>$.  Hence, the optimization of  the channels $\map E$ and $\map F$ can be reduced to the optimization of a single quantum channel $\map C :  =  \map R\circ \map E$.   

We call  the   above   scenario the {\em $U_x$-to-$V_x$ learning problem}. Its overall structure is summarised   in Figure \ref{learn_x}. 
 The temporal separation between the training phase and the execution phase makes the  $U_x$-to-$V_x$ learning problem distinct from the  problem of simulating the gate $V_x$ using the gate $U_x$ as a resource \cite{chiribella2008optimal,bisio2014optimal,chiribella2015universal,yang2017units,miyazaki2017universal,quintino2018reversing}.  In that problem, the gate $V_x$ is simulated by  an {\em arbitrary} circuit using the gate $U_x$, not necessarily a circuit of the form depicted in Figure \ref{learn_x}.

\iffalse
The $U_x$-to-$V_x$  learning problem is also related to the problem of programming unitary gates \cite{nielsen1997programmable,vidal2002storing,hillery2002probabilistic,vidal2002storing,brazier2005probabilistic,ishizaka2008asymptotic,bartlett2009quantum,sedlak2018optimal}.
  In this case, the task is to design a set of states $( |\phi_x\> )_{x\in\set X}$, called the {\em program states},  which are capable of controlling  the gates $(V_x)_{x\in\set X}$  with maximum accuracy. In this problem, no assumption is made on the functional dependence of the state $|\phi_x\>$ from the parameter $x$. In particular, it is not assumed that $|\phi_x\>$ should be of the form $|\phi_x\> =  (  U_x\otimes I_{\rm A})|\phi\>$ for some fixed state $|\phi\>$ and for a given set of unitary gates $( U_x)_{x\in\set X}$.     The $U_x$-to-$V_x$  learning problem can be regarded as a  {\em constrained quantum  programming problem}, where the program state are required to be of the specific form (\ref{learning}).   
\fi

A general result by Bisio {\em et al} concerns the case where the set $\set X$ is a group and the mappings $x\mapsto U_x$ and $x\mapsto  V_x$   (or $x\mapsto  V_x^\dag$) are two unitary representation of the group $\set X$.   In this scenario,  the authors   showed that  the optimal learning performance can be achieved with a purely classical memory \cite{bisio2010optimal}.  In this paper, we present an instance of $U_x$-to-$V_x$ learning problem that evades Bisio {\em et al} no go theorem. In our scenario,   $x$ is a rotation   $g\in  \grp {SO} (3)$, the probe is a spin-$j$ particle ${\rm P}_j$, the  training gate is  the unitary gate $U_g^{(j)}$ that implements the rotation $g$ on the probe, the target is a   qubit,   and the target gate is the rotation $V_{\theta,g} $ defined  by 
	  \begin{align}\label{Vthetag}
	    V_{\theta,g}    :=   U_g  \,  V_\theta  \,  U_g^\dag  \, ,       
    \end{align} 
    where $\theta  \in  [0,2\pi)$ is a fixed, but otherwise arbitrary   angle,   $U_g$     is the 2-by-2 unitary matrix representing the rotation $g$, and $V_\theta  = \cos \frac \theta 2  \,  I  -  i  \sin \frac \theta 2  \,  \sigma_z  $ is the 2-by-2 matrix representing  a rotation by $\theta$ about the $z$-axis. Since the rotation angle is fixed, the target operations do not form a group, and therefore our learning problem falls outside the hypotheses of Bisio {\em et al}'s no go theorem.

\iffalse    
Ideally, the machine should learn how to perform {\em all} possible rotations around  the direction $\st n   =  g \, \st e_z$, obtained from the $z$ axis $\st e_z  =  (0,0,1)$ through the action of the rotation $g$.   
 This means that all the operations in the training phase should be  independent  of the rotation angle $\theta$. %In Figure \ref{learn_x},    the auxiliary system $\rm A$, the input state $|\phi\>$, and the encoding channel $\map E$ should be independent of $\theta$.    
 However, in the following we  will not impose this condition: in order to identify the ultimate accuracy limit,   we will allow all the operations to depend on $\theta$.   Interestingly, for $j\not = 1$ we will {\em prove} that all the operations in the training phase are indeed independent of $\theta$.   
 \fi
 
 \begin{figure}[h!]
	\centering
	\includegraphics[width=0.82\textwidth]{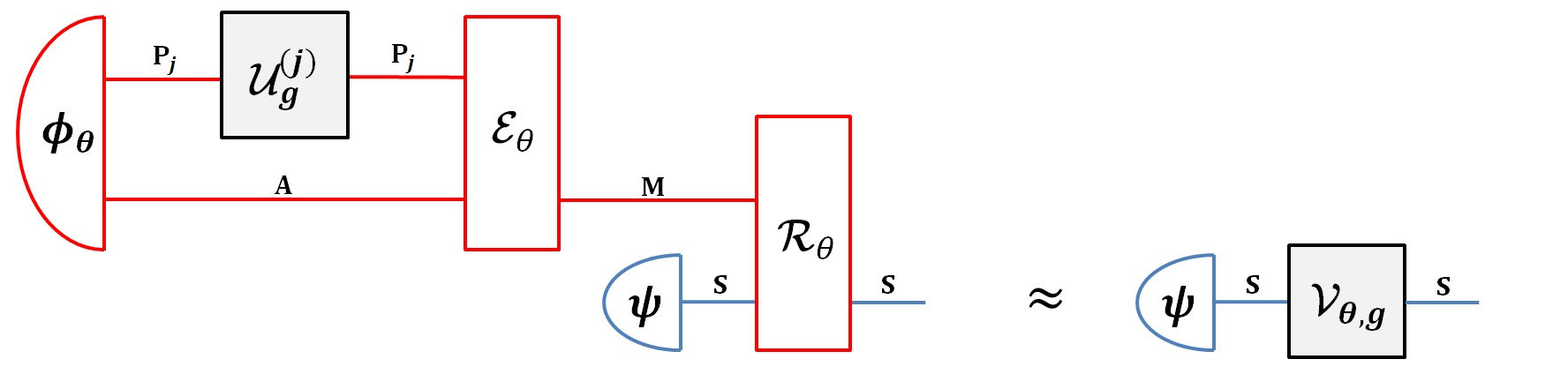}
	\caption{
		\textbf{Learning from a rotation gate.  }   {}	A machine has access to a rotation gate $U_g^{(j)}$, which implements a rotation $g$ on a quantum particle of spin-$j$, denoted by ${\rm P}_j$.   The goal is to learn how to perform rotations on a target qubit, denoted by $\rm S$. Specifically, the machine is designed to perform  the qubit gate $V_{\theta, g}  = U_g V_\theta  U_g^\dag$, where $\theta$ is a generic angle, and $V_\theta$  is a rotation by  $\theta$ about the $z$ axis.  In the training phase, the machine probes the gate $U_g^{(j)}$ by preparing the spin-$j$ particle and an auxiliary system $\rm A$ in a joint state $|\phi_\theta\>$. The output state 
		$ \big(  U_g^{(j)}  \otimes   I_{\rm A}  \big)  \,   |\phi_\theta\> $ is then stored in the internal memory of the machine and is retrieved  in the execution phase, when  the machine performs a joint operation $\map R_{\theta}$ designed to approximate the action of the target gate $V_{\theta,g}$.    }  
	\label{learn_V}
\end{figure}

The $U_g$-to-$V_{\theta,g}$ learning problem is also  relevant to the study of quantum reference frames  \cite{bartlett2007reference}.      Suppose that  two distant parties, Alice and Bob, do not share a reference frame for directions.  This means that Bob's Cartesian axes $\st n_x^{(B)},$ $\st n_y^{(B)}$, and $  \st n_{ z }^{(B)}$ are related to Alice's Cartesian axes  $\st n_x^{(A)},$ $\st n_y^{(A)}$, and $  \st n_{ z }^{(A)}$ by an unknown element of the rotation group $\grp{SO}  (3)$, namely $  \st n_i^{(B)}  =  g  \st n_i^{(A)}$, for all $i\in  \{x,y,z\}$.   Now, imagine that Bob   wants to  rotate a qubit by an angle $\theta$  about the direction of Alice's $z$-axis.  To assist Bob in this task, Alice will send him a quantum system carrying information about her reference frame. If the transmitted system is a spin-$j$ particle, prepared by Alice in the state $|\phi_\theta\>$, then Bob will receive the particle in the state $U_g^{(j)}  |\phi_\theta\>$, owing  to the mismatch of their reference frames. 
Using the state $U_g^{(j)}  |\phi_\theta\>$ as a resource, Bob can attempt to execute the desired rotation, corresponding to the unitary gate $  V_{\theta,  g}  =  U_g  V_\theta U_g^\dag$.   More generally, Alice could send Bob a spin-$j$ particle together with an auxiliary particle $\rm A$ whose state space is invariant under rotations.  In this case, Bob will receive the state $\left(   U_g^{(j)}  \otimes I_{\rm A}\right)  |\phi_\theta\>$, where $|\phi_\theta\>$ is the  initial state of the spin-$j$ and the auxiliary particle.   In this setting, the search for the optimal communication protocol between Alice and Bob is equivalent to the search of the optimal learning strategy for the  $U_g$-to-$V_{\theta,g}$ learning problem.

%	   {\em Given  black box access to a rotation gate $U_g^{(j)}$,  learn how to perform the qubit gate $V_g  = U_g V_\theta  U_g^\dag$, where $\theta\in  [0,2\pi)$ is a fixed angle and $V_\theta$  is a rotation by an angle $\theta$ about the $z$ axis (see Figure \ref{learn_V}).}
 A diagrammatic representation of the $U_g$-to-$V_{\theta,g}$ learning problem is provided in Figure \ref{learn_V}.     The spin-$j$ probe and the auxiliary system $\rm A$ start off in the state     $|\phi_\theta\>$. Then, the probe is sent through the gate $U_g^{(j)}$.  After the action of the gate $U_g^{(j)}$,  the probe  and system $\rm A$ will be in the state 
	\begin{align}\label{phithetag}
		|\phi_{\theta,g}\>    :=   \big(  U_g^{(j)}  \otimes   I_{\rm A}  \big)  \,   |\phi_\theta\>   \, ,
	\end{align}
	where $I_{\rm A}$ is the identity on the auxiliary system's Hilbert space.    Then, the state  $|\phi_{\theta,g}\>$ is encoded in the machine's memory   via a channel $\map E_\theta$.   In the execution phase, the machine will perform a quantum channel $\map R_\theta$, transforming the input state of the memory and the target into the output state of the target.

	The average fidelity for $U_g$-to-$V_{\theta, g}$  learning task is  
\begin{align}\label{Fave}
		F_2(j,\theta)  =  \int \d g \,  \int \d \psi   \,    \<  \psi|   V_{\theta, g}^\dag  \,    \map C_\theta (|\phi_{\theta,g}   \>\< \phi_{\theta,g}|  \otimes |\psi\>\<\psi|  )   \,  V_{\theta, g}  |\psi\>  \, ,    
	\end{align}   
where $\d g$ is the normalized Haar measure over the rotation group, and $\map C_\theta  :  =    \map R_\theta  \circ (\map E_\theta\otimes \map I_{\rm S})$ is the effective channel transforming states of the composite system $  {\rm P}_{j}  \otimes \rm A \otimes  \rm S$ into states of $\rm S$.  

This leads to the following optimisation problem:
\begin{prob}\label{prob:optimiseall}   Find the  auxiliary system $\rm A$, the input state $|\phi_\theta\>$, and the  channel $\map C_\theta$ that maximise the fidelity $F_2(j,\theta)$ in Equation (\ref{Fave}). 
\end{prob}

Problem \ref{prob:optimiseall} reduces to Problem \ref{prob:channelonly} of the previous subsection if  system $\rm A$ is trivial, and if the initial state $|\phi_\theta\>$ is the spin coherent   $|j,j\>$, independently of $\theta$.  
In this case, the state  (\ref{phithetag}) inputted  in the quantum machine is  the spin coherent state $U_g |j,j\>  =  |j,j\>_{\st n(g)}$, where $\st n (g)$ is the rotated $z$-axis $\st n(g):  =  g   \st e_z$, $\st e_z=  (0,0,1)$. 
 Since the rotation $g$ is chosen at random according to the Haar measure,    the direction $\st n(g)$ is distributed uniformly over the unit sphere  (see {\em e.g.} Section 4.1 of Holevo's textbook \cite{holevo2011probabilistic}).  Hence,  one has 
	\begin{align}
	\nonumber F_2(j,\theta)    &=  \int \d g \,  \int \d \psi   \,    \<  \psi|   V_{\theta, g}^\dag  \,  \map C_\theta  \Big(   |\phi_{\theta,  g}\>\<\phi_{\theta,g}|     \otimes |\psi\>\<\psi|  \Big)   \,  V_{\theta, g}  |\psi\>\\
\nonumber     &=  \int \d g \,  \int \d \psi   \,    \<  \psi|   V_{\theta, \st n (g)}^\dag  \,  \map C_\theta  \Big(   |j,j\>_{\st n(g)}\<j,j|_{ \st n(g)}      \otimes |\psi\>\<\psi|  \Big)   \,  V_{\theta, \st n( g)}  |\psi\>\\
 \nonumber &= \int \d \st n  \,     \int \d \psi   \,    \<  \psi|   V_{\theta, \st n }^\dag  \,  \map C_\theta  \Big(   |j,j\>_{\st n}\<j,j|_{\st n}      \otimes |\psi\>\<\psi|  \Big)   \,  V_{\theta, \st n}  |\psi\>\\
 & =  F_1(j,\theta)  \, .
 \end{align}
 
Hence, the fidelities   $F_1(j,\theta)$ and $ F_2(j,\theta) $ coincide when the input state $|\phi_\theta\>$ is the spin coherent state $|j,j\>$.  Under this condition,  both fidelities $F_1(j,\theta)$ and $ F_2(j,\theta) $ are maximised by the same  quantum channel $\map C_\theta$. 
%  Later in the paper, we will show that  the optimal input state $|\phi_\theta\>$ is indeed the spin coherent state   $|j,j\>$ for every value of $j$ other than $j=1$.   

	\section{Optimal quantum strategies}\label{sec:accuracysize}
	Here   we determine the optimal quantum strategies for learning rotations around an unknown direction.   We solve   Problems \ref{prob:channelonly}  and \ref{prob:optimiseall}  defined in the previous section for all values of the spin $j$ and for all values of the rotation angle $\theta$.    For  $j>1$, we show that the optimal state for Problem \ref{prob:optimiseall} is the spin coherent state $|j,j\>$, and therefore the optimal fidelity coincides with the optimal fidelity for Problem \ref{prob:channelonly}. 
	   In both problems, the best approximation of the target rotation is realised by  setting up an isotropic Heisenberg interaction between the target and the probe.
	For  $j= 1/2$ and $j=1$, we find some  curious features  of the optimal strategies. 
	%including a transition from unitary to non-unitary operations when the rotation angle approaches $\pi$.   
	Notably,  the optimal solution of Problem \ref{prob:optimiseall} deviates from the optimal solution of Problem \ref{prob:channelonly} for $j=1$ when the  rotation angle approaches $\pi$.  
	%For  $j=1/2$,   the optimal solutions of Problems \ref{prob:channelonly} and  \ref{prob:optimiseall} coincide for all values of $\theta$.
	% although the optimal strategy exhibits a  transition when $\theta$ approaches $\pi$. 
	%The deviation is characterised by  a transition of the  optimal input state from $|1,1\>$ to $|1,0\>$. 
	%the eigenstate of $J_x$ with eigenvalue $m=0$. 
	% pointing in direction  and the best approximation of the target rotation is achieved by measuring the probe and conditionally operating on the target.   Instead, for Problem \ref{prob:channelonly} the optimal operation remains unitary. 
	\subsection{Structure of the optimal solution of Problem \ref{prob:optimiseall}}
	
Here we focus on Problem \ref{prob:optimiseall} and  determine  the structure of its optimal solution. The main result is the following theorem:  

\begin{theo}\label{theo:optimalstrategy}
The optimal strategy for  learning the target gate $V_{\theta, g}  =  U_g V_\theta  U_g^\dag$ from the training gate $U_g^{(j)}$ has the following features: 
\begin{enumerate}
\item no auxiliary system is needed
\item the optimal input state is an eigenstate of $J_z$
\item the optimal quantum channel is {\em rotationally covariant}, namely 
\begin{align}\label{covariant}
		\map C_\theta   \left(  \map U_g^{(j)}  \otimes \map U_g  \right)   =    \map U_g  \map C_\theta    \qquad \forall g \in  \grp {SO} (3) \, ,
	\end{align}  
	where $\map U_g$ and $\map U_g^{(j)}$ are the quantum channels induced by the unitary gates $U_g$ and $U_g^{(j)}$, respectively.
\end{enumerate}
\end{theo}

The theorem follows from two lemmas:

	\begin{lemma}\label{lem:jm}
	No auxiliary system is needed in the optimal strategy for 
	learning the gate $V_{\theta, g}$ from the gate $U_{g}^{(j)}$.    The optimal input   is an eigenstate of the $z$-component of the angular momentum.
	\end{lemma}
	\Proof    Note the target gate $V_{\theta, g}$ satisfies the relation  $V_{\theta, g}  =  V_{\theta, gh}$ for every rotation $h$ around the $z$ axis.  Then, the fidelity (\ref{Fave}) can be rewritten as  
	\begin{align}
		\nonumber 	F_2(j,\theta)   & =  \int \d h \,  \int \d g \,  \int \d \psi   \,    \<  \psi|   V_{\theta, gh}^\dag  \,  \map C_\theta  \big( \phi_{\theta, g}    \otimes \psi   \big)   \,  V_{\theta, gh}  |\psi\>   \\
		&  = \int \d h \,   \int \d k \,  \int \d \psi   \,      \<  \psi|   V_{\theta, k}^\dag  \,  \map C_\theta  \big( \phi_{\theta, kh^{-1}}    \otimes \psi   \big)   \,  V_{\theta, k}  |\psi\>     \, ,    
	\end{align}  
	where we used the shorthand notation $\chi  :  = |\chi\>\<\chi|$, and we derived  
  the second equality  from
   the invariance of the Haar measure with the change of variables  $k=  gh$.
   		% to denote the projector on a generic vector $|\chi\>$. 
	 Defining the average state 
	\begin{align}
		\<  \phi_\theta  \>     =  \int \d h  ~  \phi_{\theta, h^{-1}}
	\end{align}
	and its rotated version $\<\phi_\theta \>_k   =  \big(U_k^{(j)} \otimes I_{\rm A}  \big  )  \, \<\phi_\theta \> \,   \big(U_k^{(j)} \otimes I_{\rm A}  \big  )^\dag$, the fidelity can be expressed as  
	\begin{align}
		F_2(j,\theta)   & =   \int \d k \,  \int \d \psi   \,    \<  \psi|   V_{\theta, k}^\dag  \,  \map C_\theta  \big(  \< \phi_\theta \>_k    \otimes \psi \big)   \,  V_{\theta, k}  |\psi\>   
	\end{align}

	Since $\<\phi_\theta\>$ is the average of $\phi$ over all rotations about the $z$ axis, it can be expressed as
	\begin{align}
	    \<  \phi_\theta\>   =     \sum_{m  =  -j}^{+j}   \,    p^{(\theta)}_{m}   \,        |j,m\>\<j,m|  \otimes  |\alpha^{(\theta)}_m\>\<\alpha^{(\theta)}_m|  \, ,  
    \end{align}
	where $\{p^{(\theta)}_m\}_{m=-j}^j$ is a probability distribution, and each $|\alpha^{(\theta)}_m\>$ is a pure state of the auxiliary system.     Since the fidelity is linear in the input state, the optimal choice is to pick one of the terms in the mixture, such as $|j,m\>\<j,m|  \otimes  |\alpha^{(\theta)}_m\>\<\alpha^{(\theta)}_m|$.   Moreover, the state of the  the auxiliary  system can be absorbed in the definition of the channel $\map C_\theta$.  This concludes the proof that the optimal input state can be chosen to be $|j,m\>$ without loss of generality and that no auxiliary system is needed. \qed 
   \medskip  
   
   Consistently with the above result,  we will omit the auxiliary system $\rm A$ from now on. 
   
	\begin{lemma}\label{lem:jm}
 The optimal channel $\map C_\theta$ for learning the gate $V_{\theta, g}$ from the gate $U_{g}^{(j)}$ can be chosen to be covariant without loss of generality. 
	\end{lemma}	
	\Proof 
	The optimality of covariant channels follows from the following chain of equalities:   
	\begin{align}
	\nonumber F_2(j,\theta)   & =  \int \d g \,  \int \d \psi   \,    \<  \psi|   U_g  V_\theta^\dag   U_g^\dag   \,  \map C_\theta  \big( \phi_{\theta, g}  \otimes \psi  \big)   \,  U_g  V_\theta  U_g^\dag  |\psi\>   \\
	\nonumber  & =  \int \d g \,  \int \d \psi'   \,    \<  \psi'|  V_\theta^\dag   U_g^\dag   \,  \map C_\theta  \big( \map U_g^{(j)}  (\phi_\theta)    \otimes \map U_g (\psi')  \big)   \,  U_g  V_\theta   |\psi'\>  \\
	  &  = \int \d \psi'   \,    \< \psi' |   \,  V_\theta^\dag   \map C'_\theta        \big (   \phi_\theta \otimes \psi'  \big)    \,  V_\theta     |\psi'\>  \, ,
	\end{align}  
  having defined $|\psi'\> := U_g^\dag  |\psi\>$ in the second equality,   and  $ \map C_\theta'  : =   \int \d g  \,     \map U_g^\dag  \map C_\theta   \big(  \map U_g^{(j)}\otimes \map U_g \big)$ in the third equality.    
  Since $\map C_\theta'$ is covariant, the above equality shows that every channel  can be replaced by  a covariant channel with exactly the same fidelity.     \qed 

\medskip   
Covariant channels have the same performance for all possible training gates. Hence, for a covariant channel $\map C_\theta$  the fidelity can be rewritten as 
  
\begin{align}\label{Fave'}
	F_2(j,\theta)  =   \int \d \psi   \,    \<  \psi|   V_\theta^\dag   \,  \map C_\theta  \big( \phi_\theta  \otimes \psi  \big)   \,   V_\theta  |\psi\>  \, .  
	\end{align}

	\subsection{Choi operator formulation}
	Theorem \ref{theo:optimalstrategy} guarantees that the optimal input state for learning the gate $V_{\theta, g}$ from the gate $U_{g}^{(j)}$   is an eigenstate of $J_z$. Let us denote it generically as $|j,m_\theta\>$, for some $m_\theta$ between $-j$ and $+j$, possibly depending on the rotation angle $\theta$.  
	  In the following we will search for the optimal value  $m_\theta$ and for the optimal covariant channel $\map C_\theta$.  

	First of all, we rewrite the average fidelity as
    \begin{align}\label{horodecki}   
    	 F_2(j,\theta)  =\dfrac{1}{3}+\dfrac{2}{3} \, F_2^{\rm (e)}(j,\theta)  \, , 
    \end{align}
	where $F_2^{\rm (e)}$ is the   entanglement fidelity  \cite{horodecki1999general}, defined as 
    \begin{align}
    	F_2^{\rm (e)}(j,\theta)    :=     \<\Phi^+ |    ( V_\theta \otimes I_{\rm R})^\dag \,\left[   \left(  \map C_\theta  \otimes \map I_{\rm R}\right)  \left(  |j,m_\theta\>\<j,m_\theta| \otimes   \Phi^+   \right)\right]  \,     (  V_\theta\otimes I_{\rm R}) |\Phi^+\> \, ,
    \end{align}
       $|\Phi^+\>     =  (|0\>\otimes |0\>   +  |1\>\otimes |1\>)/\sqrt 2 $ being the canonical maximally entangled state and $\rm R$ denoting a reference qubit, entangled with the target qubit.   In turn, the entanglement fidelity can be expressed as 
	\begin{align}\label{Feave-change}
       	F_2^{\rm (e)}(j,\theta)
       	&= \frac 12    ~  \big( \<j,m_\theta| \otimes  \<   \Phi_{\theta}| \big) \, C_{\theta}  \, \big(|j,m_\theta\> \otimes |\Phi_{\theta}\>\big) \, ,
    \end{align}
	where  $ |\Phi_\theta\>$ is the rotated maximally entangled state 
	\begin{align}\label{phitheta}     
		|\Phi_\theta\>  =  (   V_\theta \otimes  I_{\rm R})  |\Phi^+\>  \, ,
	\end{align} and $C_\theta$ is  the Choi operator 
	\cite{choi1975completely}
    \begin{align}
	    C_\theta =  2 (2j+1)  \,  (\map I_{{\rm R}_j} \otimes  \map C_\theta \otimes    \map I_{\rm R}   )   \left(  \Phi^+_j  \otimes  \Phi^+ \right) \, ,
    \end{align}
     ${\rm R}_j$  being  a reference system of dimension $2j+1$,    $\map I_{{\rm R}_j}$  ($\map I_{\rm R}$) being the identity map on the reference system ${\rm R}_j$  (${\rm R}$),  and $  |\Phi^+_j\> =  \sum_{m}  \,  |j,m\>\otimes |j,m\>/\sqrt{2j+1}$ being the canonical maximally entangled state in dimension $2j+1$.   
    
The problem is  to  maximise  the fidelity    (\ref{Feave-change}) over all Choi operators of covariant channels.   The set of the possible Choi operators is characterised by the following three conditions: 
	\begin{enumerate}
	\item {\em Covariance \cite{chiribella2009optimal}:}   $[  C_\theta   ,    \overline U_g^{(j)}  \otimes U_g  \otimes  \overline U_g ]=  0$ for all rotations $g\in\grp{SO}  (3)$  (here $   \overline U^{(j)}_g$ and $\overline U_g$  denote the entry-wise complex conjugates of the matrices $U_g^{(j)}$ and $  U_g$, respectively.)
		\item {\em Positivity:}   $C_\theta$ is positive semidefinite, denoted as $C_\theta  \ge 0$
		\item  {\em Trace preservation:}   $\Tr_{\rm out}  [  C_\theta]  =  I_{\rm in}$, where $\Tr_{\rm out}$ denotes the trace over the output,  and $I_{\rm in}$ denotes the identity over the input. 
		% and $d_{\rm in}  =  2 (2j+1)$ is the dimension of the input space.
		 
	\end{enumerate}

	We now put the above conditions in a form that is convenient for optimization.  

	{\em Covariance.}   The covariance condition can be further simplified using the fact that complex conjugate representations of the rotation group are unitarily equivalent.    Defining the operator
    \begin{align}
	    C^*_{\theta} := \left(  e^{-i\pi  J_y}\otimes I\otimes\sigma_{y} \right) \ C_{\theta} \ \left( e^{i\pi  J_y}\otimes I\otimes\sigma_{y}\right) \ . 
    \end{align}
	the covariance condition becomes 
	\begin{align}\label{cov}
        \left[C^*_{\theta}, U_{g}^{(j)}\otimes U_{g}\otimes U_{g}\right]=0,\qquad \forall g\in \grp{SO}(3)\, .
    \end{align}   
	At this point, the total Hilbert space can be decomposed into orthogonal subspaces, corresponding to different values of the total angular momentum.    Specifically, the angular momentum  takes values $j-1, j,$ and  $j+1$, and the total Hilbert space is decomposed as 
    \begin{align}\label{subspaces}
	    \C^{2j+1} \otimes \C^2 \otimes \C^2  =  \C^{2j-1}   \oplus \C^{2j+3}   \oplus  \left(    \C^{2j+1}  \otimes \C^2 \right) \, . 
    \end{align}        
	Relative to this decomposition, using Schur’s	lemmas and the covariance condition  (\ref{cov}), the operator $C_\theta^*$ can be written  as:
 	\begin{align}\label{Choi}
      	C^*_{\theta}=
	\alpha P_{j+1}\oplus\beta P_{j-1}\oplus  \left(  P_{j}\otimes M  \right)\ ,
    \end{align}
    where $P_l$ is the projection on the factor with total angular momentum $l$,  $\alpha$ and $ \beta$ are complex coefficients, and  $M$ is a complex 2-by-2 matrix.    
    
    {\em Positivity.}  The positivity of the operator $C_\theta$ is equivalent to the positivity of the coefficients  $\alpha, \beta$ and of the matrix $M$.  
     
	{\em Trace preservation.}  The condition of trace preservation can be conveniently expressed in terms of the real coefficients $\alpha, \beta$ and of the complex matrix $M$.  Indeed,    tracing over the output, we obtain  
    \begin{align}\label{trace}
	    \Tr_{\rm out}  [ C_\theta^* ]   =  \alpha  \,     \frac{2j+3}{2j+2} \,  P_{j+\frac 12}   +  \beta  \,       \frac{2j-1}{2j} \,  P_{j-\frac 12}  +   \<+|  M  |+\>       \frac{2j+1}{2j+2} \,  P_{j+\frac 12}   +   \<-|  M  |-\>       \frac{2j+1}{2j} \,  P_{j-\frac 12}   \, , 
    \end{align} 
    for a suitable choice of basis   $\{|+\>,  |-\>\}$. 
    Using Eq. (\ref{trace}), the trace preservation condition $\Tr_{\rm out}  [  C_\theta]  =  I_{\rm in}$ becomes 
    \begin{align}\label{Choi-coefficient'}
	    \begin{cases}
    	\dfrac{2j+3}{2j+2}  \, \alpha+\dfrac{2j+1}{2j+2} \,   \<  + |  M|+\>  =1\\  \\ 
    	\dfrac{2j-1}{2j}\,  \beta+\dfrac{2j+1}{2j}  \,      \<  - |  M|-\>  =1    \, .      	\end{cases}
    \end{align}

	{\em Figure of merit.}
    In terms of the operator $C_\theta^*$, the entanglement fidelity can be expressed as  
    \begin{align}
       	F_2^{\rm (e)}(j,\theta)
       	&= \frac 12     ~ \big(  \<j,-m_\theta| \otimes   \<   \Phi^*_{\theta}|  \big)  \, C^*_{\theta}  \,\big ( |j,-m_\theta\> \otimes |\Phi^*_{\theta}\> \big)  \, ,  \qquad {\rm with} \qquad  |\Phi^*_\theta\>   =    (  I\otimes  -i \sigma_y) \, |\Phi_\theta\>  
    \end{align}
	%and has to be maximized over all coefficients $\alpha  \ge 0, \beta\ge 0$ and all matrices $M\ge 0$ satisfying the constraints (\ref{Choi-coefficient'}).   
	The expression can be further simplified by decomposing the state $  |j, -m_\theta\>\otimes |\Phi^*_{\theta}\>$ on the subspaces of Eq. (\ref{subspaces}).   After a bit of labor with the Clebsch-Gordan coefficients, we  find the decomposition  
	\begin{align}
		|j,-m_\theta\>  \otimes |\Phi^*_{\theta}\>  =   a \,  |j+1,-m_\theta\>   +    b\,  |j-1, -m_\theta\>  +  c_+  \,  |j,-m_\theta\> \otimes   |+\> + c_-  \,  |j,-m_\theta\>\otimes  |-\>   \, ,      
    \end{align}
	with 
	\begin{align}
		\nonumber a  =  -i\sin \frac \theta 2  \,  \sqrt{\frac{  (j+1+m_\theta)(j+1-m_\theta)}{(j+1)(2j+1)}}   \qquad  & b  =+i  \sin \frac \theta 2  \,  \sqrt{\frac{  (j+m_\theta)(j-m_\theta)}{j(2j+1)}}  \\
		c_+ =   -  \cos \frac\theta 2 \,  \sqrt{\frac{j+1}{2j+1}}   -  i \sin \frac \theta 2  \, \frac{m_\theta}{\sqrt{  (j+1)(2j+1)}}   \qquad   &  c_- =     \cos \frac\theta 2 \,  \sqrt{\frac{j}{2j+1}}   -  i \sin \frac \theta 2  \, \frac{m_\theta}{\sqrt{  j(2j+1)}} \, .
	\end{align}
	Using the above decomposition, the entanglement fidelity can be expressed as 
	\begin{align}\label{aaabbb}
       	F_2^{\rm (e)}(j,\theta)     =  \frac{\alpha \, |a|^2  +  \beta \,  |b|^2    +    \<c|  M|c\> }2 \, , \qquad  {\rm with}  \quad  |c\>  =     c_+ |+\> + c_-  |-\> \, ,   
	\end{align}
	to be maximized over all positive coefficients $\alpha$ and $\beta$, and over all non-negative matrices $M$ satisfying the constraint (\ref{Choi-coefficient'}). 

\begin{lemma} The matrix $M$ can be chosen to be rank-one without loss of generality, namely $M  =  |v\>\<v|$ for some suitable vector $|v\>   =    v_+ |+\> +  v_- |-\>  \in \C^2$. 
\end{lemma}

\Proof The entanglement fidelity depends on the  matrix $M$ through the matrix element $\<c|  M|c\>$.  Now, one has the chain of inequalities
\begin{align}
\nonumber \<c|  M|c\> 
 &\le  |c_+|^2  \, \<+ |M|+\>  +    |c_-|^2 \,  \<-|  M| -\>  + 2 |c_+ |\, |  c_-|  \,  |  \<+  | M|-\>|  \\
\nonumber &\le  |c_+|^2   \, \<+ |M|+\>  +    |c_-|^2   \,  \<-|  M| -\>    + 2  |c_+ |\, |  c_-|   \,    \sqrt{    \<+|  M| +\>  \,  \<-|  M| -\>}\\
 &  =  \Big(  |v_+| \,  \sqrt{ \<+|  M| +\>}  +   |v_-| \, \sqrt{  \<-|  M| -\>} \Big)^2 \, , \label{poi}
\end{align}
the second  inequality following from the fact that $M$ is positive.

The first inequality holds with the equality sign when the phase of the complex number $\<  +  |  M  |-\>$ is equal to the phase of the complex number $\overline c_+  c_-$.   The  second inequality holds with the equality sign if $M$ is rank-one.   In particular, the upper bound is attained by the rank-one matrix $M'  =  |v\>\<v|$ with  $v_+   =  \sqrt{\<+ |M|+\>}$ and $v_-  =  \sqrt{\<- |M|-\>}    \overline c_+  c_-/  |c_+c_-|$. 

Since the normalization constraint  (\ref{Choi-coefficient'}) involves only the diagonal matrix elements of $M$, the  matrix $M$ can be replaced by the matrix $M'$ without loss of generality. \qed 

\medskip 

The proof of the above lemma shows that the optimal entanglement fidelity has the form

	\begin{align}\label{fidfid}
		F_2^{\rm (e)}(j,\theta)     =  \frac{ \alpha \, |a|^2  + \beta \,  |b|^2    +      \big(  |v_+| \,  |c_+|  +   |v_-| \, |c_- | \big)^2}2   \, ,   
	\end{align}
	with $|v_\pm|  =  \sqrt{\<\pm  |  M |\pm \>}$.  The maximum of the fidelity (\ref{fidfid}) under the constraints  (\ref{Choi-coefficient'}) can be determined with the  method of Lagrange multipliers. In the following we present the  result of the maximization, leaving the details to Appendix \ref{app:lagrange}.

	\subsection{Optimal quantum strategy for $j>1$}
	For $j>1$, it turns out that Problems \ref{prob:channelonly} and \ref{prob:optimiseall} have the same optimal solution:
		\begin{theo}
		When $j>1$, the optimal  probe  state   for learning the gate $V_{\theta,g} = U_g V_\theta  U_g^\dag$ from the gate $U_g^{(j)}$ is $|j,j\>$  for every value of $\theta$.   For both Problems \ref{prob:channelonly} and \ref{prob:optimiseall}, 
		%optimal entanglement fidelity is 
		%		\begin{align}\label{FidOptimal}
		%		F^{(\rm e)}_{\rm opt}(j ,\theta)=    \frac{1 + \sqrt{ 1  +    \frac{2j+1}{j^2}  \, \left (\cos \frac \theta 2 \right)^2  }   +   \frac{2j+1}{2j^2}  \, \left (\cos \frac \theta 2 \right)^2   }{2 \, (1+ \frac 1{2j})^2}  \, .
		%	\end{align}  
		%	Equivalently, the 
		optimal average   fidelity over all pure input states is
		\begin{align}\label{Foptnew}
			F_{\rm opt}(j ,\theta)=  \frac{1}{3} + \frac{1 + \sqrt{ 1  +    \frac{2j+1}{j^2}  \, \left (\cos \frac \theta 2 \right)^2  }   +   \frac{2j+1}{2 j^2}  \, \left (\cos \frac \theta 2 \right)^2   }{3 \, (1+ \frac 1{2j})^2} \, ,
		\end{align} 
		and has the asymptotic expression 
		\begin{align}\label{Foptapp}
			F_{\rm opt}(j ,\theta)   &   =  1-\dfrac{1-\cos \theta }{3j}+
			%\dfrac{7-6\cos\theta-\cos^{2}\theta}{24j^{2}}+
			O\left(\dfrac{1}{j^2}\right)  \, .
		\end{align}
	\end{theo}

	The optimality of the probe state $|j,j\>$ is in agreement with a result by Holevo on the optimal estimation of directions, cf. Section  4.10 of \cite{holevo2011probabilistic}. In other words, the optimal probe state for  learning how to rotate about an unknown direction coincides with the optimal probe state for producing a classical estimate of such direction, as long as $j$ is larger than 1.  It is worth stressing, however, that the optimal quantum   strategy for  rotating about an unknown direction is {\em not} based on estimation: in Section  \ref{sec:benchmark} we will show that no estimation-based strategy can achieve the optimal quantum fidelity  (\ref{Foptnew}).

	\begin{figure}[ht]
		\centering
		\includegraphics[width=0.7\textwidth]{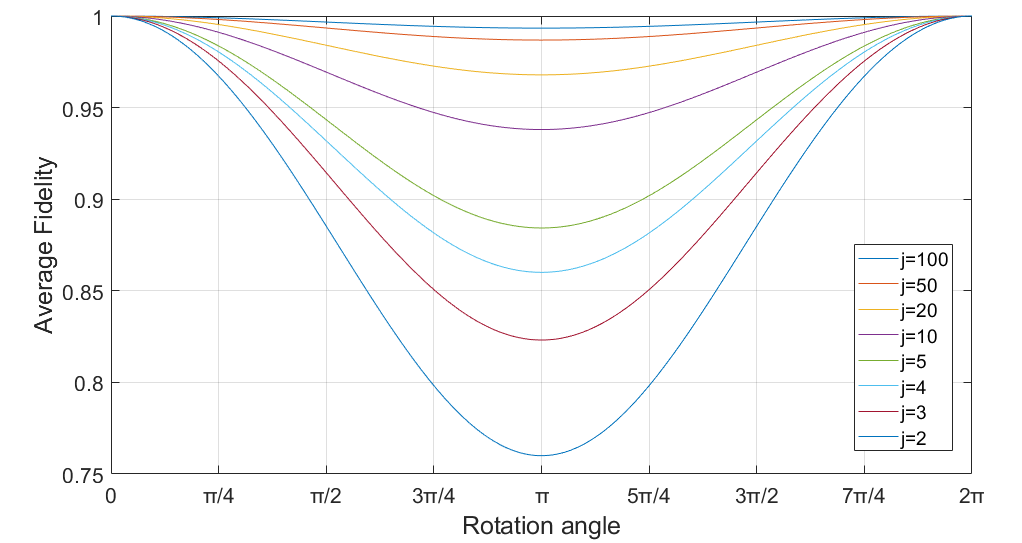}
		\caption{\footnotesize
		\textbf{Optimal average fidelity for $j>1$. }     The dependence of the fidelity on the rotation angle $\theta$ is illustrated for different values of the  spin from $j=2$ to $j=100$.   The fidelity is minimum for $\theta  =  \pi$. 
		}
		\label{fig2}  
	\end{figure}
	The exact values of the average fidelity are plotted in Figure \ref{fig2} for various values of $j$ from $j=2$ to $j=100$.  
 	Note that the fidelity decreases monotonically with the rotation angle $\theta$. Intuitively, rotating by smaller angles is easier, because the uncertainty about the rotation axis has less influence on the performance. The easiest rotation is the identity $(\theta=0)$, which is independent of the rotation axis and therefore can be implemented without error.    The hardest rotation is the spin flip, corresponding to $\theta =\pi$.  In this case, the average fidelity has the simple form  
	\begin{align}\label{Fidpi}
		F_{\rm opt}(j ,\pi)  = 1    -  \dfrac{ 8j +2}{12j^2 +12j + 3}  
		  \, .
	\end{align}

 Note that, since the optimal probe state is $|j,j\>$,  the optimal channel $\map C_\theta$  for  Problem \ref{prob:channelonly} coincides with the optimal channel $\map C_\theta$ for Problem  \ref{prob:optimiseall}. 
		In Appendix \ref{app:physical}, we show that an optimal channel $\map C_\theta$  can be attained by setting up  an isotropic  Heisenberg  interaction between the memory spin and the target spin.
	Explicitly, we show that the maximum fidelity (\ref{Foptnew}) is attained by the channel 
	\begin{align}\label{map_rel_C_theta36}
		\map C_{\theta, \rm Hei}  (\rho)   =     \Tr_{{\rm P}_j}  \big[   U_\theta    \rho  U_\theta^\dag \big] \, ,  
	\end{align}
	where $\Tr_{{\rm P}_j}$ denotes the partial trace over the probe, and $U_\theta$ is the unitary operator
    \begin{align}\label{optimalU}
        U_\theta  =     \exp  \left[    -i f(\theta) \, \frac{  \,   \st J \cdot \bs \sigma}{2j+1}  \right]  %P_{j +  \frac 12}  + e^{i f(\theta)} \,  \, P_{j- \frac 12}    
    \end{align}
	in which       $\bs \sigma  =  (\sigma_x, \sigma_y,  \sigma_z)$ is the vector of the three Pauli matrices,   $\st J \cdot \bs \sigma   = \sum_{i=x,y,z} J_i\otimes  \sigma_i$   is the Heisenberg coupling, and   $ f(\theta)$ is the function 
    %  =  [ \theta  -  { \sin \theta}/(2j) ]/(2j+1)$ at the leading order.
   \begin{align}\label{ftheta}
		f(\theta) = \arccot  \left[  \cot \theta    +    \dfrac{1}{(2j+1)\sin\theta}\right]  +  s(\theta)
	\end{align}
	where $s(\theta)  =  0$ for $\theta \in  [0,\pi]$, and $s(\theta)  =  \pi$ for $\theta  \in  (\pi,2\pi)$. 
    %\begin{align}\label{ftheta_old}
    %    f(\theta)  &= \theta_+  +  \theta_-   =  \arctan  \frac {\sin\theta}{\cos\theta  +  \frac 1 {2j+1}} \, .
    %\end{align}
    Note that $f(\theta)$ is approximately equal to $\theta$ in the large $j$ limit.

	Physically, the unitary evolution  (\ref{optimalU}) can be realized by setting up an isotropic Heisenberg  interaction, described by the Hamiltonian $H  =   \alpha  \, \st J \cdot \bs \sigma$, for some suitable coupling constant $\alpha$, and by letting the two spins evolve for  time 
    \begin{align}
	    t (\theta)  =       \frac{f(\theta)}{(2j+1)  \alpha \hbar}\, ,
    \end{align}
    depending on the  angle  $\theta$ of the target rotation.       Remarkably,  the same probe states and the same interaction can be used to control the full time evolution of the target system:  one has only to adjust  the interaction time [determined by  the angle $f(\theta)$] based on the evolution time in the target dynamics [determined by the angle $\theta$].  For example, we can set $\theta   =  \omega  t$ and simulate the    precession of a spin-$1/2$ particle around the direction indicated by the memory state.

 An important feature of the optimal strategy is that the  optimal probe state is independent of the rotation angle $\theta$.  Since the operation of storing the state $U_g | j,j\>$ in the quantum memory is also independent of $\theta$, it follows that all the operations in the training phase can be accomplished without knowing the rotation angle.  This offers the possibility to decide the value of $\theta$ at later times.   In fact,   the machine can optimally approximate  the full continuous-time dynamics of the target particle, because the optimal operations for different $\theta$ corresponds to unitary evolutions with the same Hamiltonian, just with different evolution times.

%	It is worth stressing that the unitary realization  (\ref{optimalU})   is \emph{economical} \cite{fuchs1997optimal, niu1999two, durt2005economical}, in the sense  that it only requires an interaction between the memory and target spins, without the addition of auxiliary systems.  

 	The optimality of the Heisenberg interaction is not limited to the average fidelity. In terms of scaling with $j$,  the unitary gate (\ref{optimalU}) is optimal also for the {\em worst-case fidelity}, defined as      
	\begin{align}\label{Fwc-app}
	    F_{w}(j,\theta) =  \min_{g}  \,  \min_\psi 
	    \,  F(j,\theta, g, \psi) \, ,
    \end{align}
	where $ F(j,\theta, g, \psi)$ is the fidelity for the simulation of $V_g$ on the specific input state $|\psi\>$.  Indeed, in Appendix \ref{app:worstcase}, we show that the worst-case fidelity of the unitary gate (\ref{optimalU})   is   
	\begin{align}\label{Fwc}
        F_{\rm w, Hei} (j,\theta) = 1-\dfrac{1-\cos\theta}{j} 
        %\dfrac{\cos^{2}\theta -4\cos\theta+3}{3j^{2}}
        +O\left(\dfrac{1}{j^2}\right) \, .
    \end{align}
	Hence, the worst-case infidelity $1-   F_{\rm w, He} (j,\theta)$ has the scaling $1/j$. This  is the best  scaling one can hope for, because the  average infidelity  cannot vanish faster than   $1/j$ [as shown by Eq. (\ref{Foptapp})], and   the average  infidelity  is a lower bound to the worst-case infidelity. 

The optimality of the Heisenberg interaction answers in the affirmative a question raised by Marvian and Mann  \cite{marvian2008building}, who  assumed the Heisenberg interaction  and showed that it can be used to approximate arbitrary rotations in the limit of large $j$ limit. In the conclusion of their work, Marvian and Mann asked whether the Heisenberg interaction achieves the best scaling of the error with the spin size. Our results provide an affirmative answer, showing that the Heisenberg interaction maximizes  the average fidelity and has the optimal error scaling $O(1/j)$ in the worst-case scenario.

\iffalse	Knowing that the worst-case fidelity vanishes as $1/j$, we can estimate the error also in   scenarios where  many  gates are combined together.  For example, a quantum machine could learn how to perform $k$ single-qubit gates $(V_{\theta_1,  \st n_1},  V_{\theta_2,  \st n_2} ,\dots,  V_{\theta_k, \st n_k}  )$, for the purpose of simulating a quantum circuit,   such as   
	\begin{align}
 W_k  \, \big(  V{\theta_k,  \st n_k}  \otimes I_{\overline Q_k}\big )         \cdots  W_2 \, \big(  V{\theta_2,  \st n_2}  \otimes I_{\overline Q_2}\big) \,     W_1 \,   \big(  V{\theta_1,  \st n_1}  \otimes I_{\overline Q_1}\big) W_0 	 \, ,
	\end{align} 
	where $(W_0, W_1 \dots,  W_k)$ are fixed unitary gates acting on a multiqubit register, and  the gate $V_{\theta_i, \st n_i}$ acts on the qubit $Q_i$, and  $I_{\overline  Q_i}$ denotes the identity on the remaining qubits.     The deviation of this circuit from its ideal functionality can be evaluated by the trace distance between its output and the ideal output.  Using the fact that the trace distance is upper bounded by twice the square root of the infidelity, we obtain that  whose deviation from the ideal functionality will be bounded by a term of  size $k/\sqrt j$.   This means that, in order to have a negligible error on a circuit of $k$ gates, the machine should allocate  a memory of size $j  =  \Theta  (k^2)$ to each gate in the assumption that each     
\fi

 \subsection{Optimal quantum strategy for $j = 1/2$} \label{sec:3.5}
  	For $j=1/2$, the optimal probe state for Problem  \ref{prob:optimiseall} is still the coherent state $|j,j\>$ for every rotation angle $\theta$, and the optimal solutions of Problems \ref{prob:channelonly} and \ref{prob:optimiseall} still coincide.

Curiously, the optimal  learning  strategy exhibits a  transition  when the rotation angle approaches $\pi$.  For $|\theta-\pi| > \delta_{1/2}=  \arccos [( 4 +\sqrt 7)/9] $, the optimal fidelity is still given by Equation (\ref{Foptnew}), %which now reads, 
%\begin{align}
			%F_{\rm opt}\left(j=\frac 12 ,\theta\right)=  \frac{1}{3}   \left[   \frac 54 + \sqrt{ 1  +     \frac 14   \, \left (\cos \frac \theta 2 \right)^2  }   +     \, \left (\cos \frac \theta 2 \right)^2 \right]  \, ,
%		\end{align} 
and the optimal channel $\map C_\theta$ is still  given by Equation (\ref{map_rel_C_theta36}).

   For $|\theta-\pi|  \le \delta_{1/2}$, instead,  the optimal fidelity becomes
  	\begin{align}\label{crazy}
	 	F_{\rm opt}\left(j=\frac 12,\theta\right)  =  \frac {5-\cos\theta}{12}  - \frac {1-\cos\theta}{36(1+ 2\cos \theta)}
	\end{align}

	%For  $\theta =  \pi$, the  fidelity is $1/3$, which is higher than the $1/4$  fidelity achieved by the Heisenberg interaction (cf. Equation (\ref{Fidpi})  with $j=1/2$). 
	and  is achieved by the following strategy: 
	 \begin{enumerate}
	\item Perform a joint  measurement on the memory and the target. The measurement has  two  outcomes and is described by the  quantum operations $\map M_{\rm yes}  (\cdot)  =   M_{\rm yes}\cdot M_{\rm yes}^\dag   $   and $\map M_{\rm no}  (\cdot)   = M_{\rm no}\cdot M_{\rm no}^\dag $, with 
	\begin{align}
		M_{\rm yes}  :=  \sqrt{1-  \frac 43\,  \alpha}  \,    P_1  +  P_0   
		 \qquad {\rm and}  \qquad  M_{\rm no}   :=  \sqrt{\frac 43  \,\alpha}  P_1  \, , 
	\end{align}
	$P_l$ being the projector on the subspace with total angular momentum $l$, with $l\in \{0,1\}$. 
	\item If the measurement yields  outcome  ``$\rm yes$'', then apply the unitary gate (\ref{optimalU}), corresponding to the Heisenberg interaction, and discard the memory.    If the measurement yields outcome ``$\rm no$'',  then perform the optimal  2-to-1 universal NOT channel \cite{buvzek1999optimal}, namely the channel $\map C_{\rm UNOT}$ defined by    
	\begin{align}
	\map C_{\rm UNOT}  (\rho)    :=   \int \d g  \,   3\,    (\< 0|  \otimes  \<0|)    U_g^{\dag \otimes 2}  \,  \rho     \,  U_g^{\otimes 2}  (|0\>  \otimes |0\>)   ~  U_g  |1\>\<1|U_g^\dag   \, .  
	\end{align}
	\end{enumerate}
%The above strategy is rather counterintuitive.  
% {\em A priori}, it is not clear why the two input spins should be projected into the subspace with total spin equal to $1$.   And after the projection takes place, it is not clear why the  universal NOT gate should be the optimal conditional operation. These results are non-trivial consequences of the mathematical derivation  of the optimal fidelity  (\ref{crazy}). 
   The probability of the outcome  ``$\rm no$", corresponding to 
    the universal NOT, depends on  the  parameter $\alpha$ in Eq. (\ref{47}).   At the critical distance $|\theta-\pi| =  \arccos [( 4 +\sqrt 7)/9] $, one has $\alpha=  0$, and the optimal strategy is  realized through the Heisenberg interaction.  As the rotation angle gets closer to $\pi$,  the coefficient $\alpha$ increases, reaching its maximum value $\alpha=  2/3$ for $\theta=  \pi$.  At this point, the weight of the universal NOT is maximum. Notably, the value $\alpha=1$ is never reached, meaning that the optimal joint measurement on the input qubits is never projective.

	\subsection{Optimal quantum strategies for $j = 1$} \label{sec:3.5}
  The $j=1$ case is the only case where  Problems  \ref{prob:channelonly} and \ref{prob:optimiseall} yield different solutions.   The difference appears  when the rotation angle is within a critical distance $\delta_{1}=0.23   \pi$ from $\pi$.

For $|\pi  -\theta| > \delta_{1}$, the optimal probe state for Problem \ref{prob:optimiseall}  is  $|1,1\>$,  and therefore the   optimal solutions for  Problems \ref{prob:channelonly} and \ref{prob:optimiseall} still coincide.  The optimal average fidelity is still given by Equation (\ref{Foptnew}) and  the optimal channel $\map C_\theta$ is still  given by Equation (\ref{map_rel_C_theta36}).	
	   
For $|\theta-\pi|\le \delta_{1}$,  the optimal average fidelity for  Problem \ref{prob:channelonly} is
\begin{align}\label{Foptnew1}
			F_{1, \rm opt}(j =1,\theta)=  \frac{1}{9}    \left[ \frac {13}3 + \frac 43  \sqrt{ 1  +   3  \, \left (\cos \frac \theta 2 \right)^2  }   +   2  \, \left (\cos \frac \theta 2 \right)^2   \right]   \, ,
		\end{align} 
		corresponding to  Equation (\ref{Foptnew}) with $j=1$. The  optimal channel $\map C_\theta$ is still given by Equation (\ref{map_rel_C_theta36}).

Instead,  the optimal fidelity for Problem \ref{prob:optimiseall} is 
	\begin{align}\label{Fid1ave}
		F_{\rm 2, opt}(j=1,\theta)     =  \frac 13  + \frac 25 \,  \left(\sin \frac \theta 2\right)^2   \, , 
	\end{align}
and is attained with the probe state  $|1,0\>$,  the $p$-orbital aligned in the direction of the $z$-axis. 
	In Subsection \ref{subsect:MOj=1}, we will show that the optimal quantum fidelity (\ref{Fid1ave}) is  achievable with a purely classical memory.  Specifically, we will see that the optimal strategy is to perform a projective measurement on the probe, with the three measurement outcomes corresponding to the three Cartesian axes. The measurement outcome is then stored into  a classical memory of 2  bits.  In the execution phase, the machine rotates  the target qubit  by an angle $\pi$ about the axis corresponding to the measurement outcome.

\subsection{Optimal fidelities for $j=1/2$ and $j=1$}

\begin{figure}[ht]
	\centering
	\includegraphics[width=0.68\textwidth]{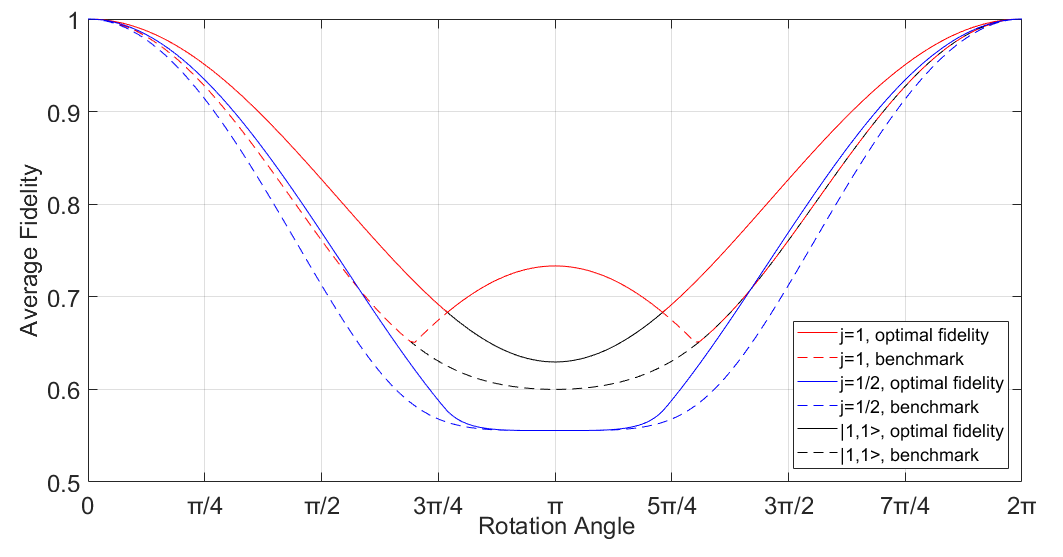}
	\caption{
		\textbf{Optimal quantum fidelities and benchmarks for $j=1/2$ and $j=1$.}     Solid curves show the maximum  of the fidelity over all quantum machines, while dashed curves provide the corresponding benchmarks, equal to the maximum fidelity over all machines equipped with a purely classical memory (derivation provided in the next Section).   For $j=1/2$, the optimal strategies for Problems \ref{prob:channelonly} and \ref{prob:optimiseall} coincide.  The fidelity of the optimal quantum strategy is  higher than the benchmark   (blue dashed line) for all values of $\theta$ except  $\theta  =  0$ and $\theta=  \pi$ (although the difference in the transition region is so small that cannot be read out from the plot).  A transition in the optimal quantum channel $\map C_\theta$  occurs at the critical distance $|\theta-\pi |=  \delta_{1/2}  :=  \arccos [( 4 +\sqrt 7)/9]  \approx 0.236\pi$. For $j=1$, the optimal strategies for Problems \ref{prob:channelonly} and \ref{prob:optimiseall} coincide for $|\theta-\pi| > \delta_{1}  \approx 0.23 \pi  $, but  become different for  $|\theta-\pi|<\delta_{1}  \approx 0.23 \pi $.      	The optimal fidelity for  Problem \ref{prob:channelonly}  (black solid curve)   is  higher than the benchmark for every $\theta \not = 0$  (black dashed line).  The optimal fidelity for Problem \ref{prob:optimiseall} (red solid curve) deviates from the optimal fidelity for  Problem  \ref{prob:channelonly}  when the distance $|\theta-\pi|$ goes below the critical value   $\delta_{1}  \approx 0.23 \pi  $.      At the critical distance, the optimal input state changes discontinuously from $|1,1\>$ to  $|1,0\>$.   In this region, the optimal  quantum fidelity becomes equal to the benchmark  (red dashed curve).  }
	\label{fig:spin12}
\end{figure}

The dependence of the fidelity on the rotation angle is plotted in Figure \ref{fig:spin12} for  $j=1$ and $j=1/2$.  The value of the optimal quantum fidelity is  contrasted with   the maximum fidelity achievable with a purely classical memory, which will be derived in Section \ref{sec:benchmark}. 
	
	\section{The quantum benchmark}\label{sec:benchmark} 
	In this section we derive the maximum fidelity achievable by learning machines with a purely classical memory of arbitrarily large size. Such fidelity  provides a benchmark that can be used to certify the experimental demonstration of quantum-enhanced learning.  
	We consider the two learning tasks corresponding to  Problem \ref{prob:channelonly} (learning from a spin coherent state) and  Problem \ref{prob:optimiseall} (learning from a rotation gate) coincide. The quantum benchmarks for these two problems coincide for all values of $j$ except $j=1$.  For $j=1$,  the two benchmarks become different when the desired rotation angle approaches $\pi$. 

%Quantum benchmarks have been extensively studied for transformations of quantum states,  such as teleportation \cite{hammerer2005quantum,adesso2008quantum,owari2008squeezing,calsamiglia2009phase}, amplification \cite{namiki2008fidelity,chiribella2013optimal,chiribella2014quantum}, and  cloning \cite{bruss1999optimal,yang2014certifying}.   In contrast, no quantum benchmark for the transformation of quantum gates has been derived so far, partly because of the technical difficulty of optimizing over the choice of conditional quantum channels, rather than conditional state preparations. The techniques developed in this section can be generally applied to quantum benchmarks for the execution of reversible qubit gates.  

	\subsection{Measure-and-operate  (MO) channels} 
	Here we consider  learning strategies where the memory $\rm M$   in Figures \ref{learn_Thermal}  and \ref{learn_V} is purely classical.   
 	In this case, the transfer of information  from the probe to the memory is described by a quantum-to-classical channel $\map E_\theta$,	 of the form  
	\begin{align}
		\map E_\theta  (\cdot)    = \sum_{y\in\set Y}  \,       \Tr[ P_{\theta, y} \,   \cdot ]\, |y\>\<y|  \, ,	
	\end{align}
	 where  $\{  |y\>\}_{y\in\set Y}$ is a set of orthogonal states of the memory,   and  $(P_{\theta, y})_{y\in\set Y}$ is a Positive Operator-Valued Measure (POVM), describing a quantum measurement on system ${\rm P}_j$  in the case of Figure \ref{learn_Thermal}, or a quantum measurement on system ${\rm P_j} \otimes \rm A$ in the case of Figure \ref{learn_V}.  
	  
	 The execution phase  consists in reading out the index $y$ from  the classical memory and performing a conditional operation $\map O_{\theta, y}$ on the system.      Hence, the channel $\map R_\theta$ has the form 
	 \begin{align}
		 \map R_\theta  (\cdot)  =    \sum_y   \map O_{\theta, y}   \Big(  \Tr_{ \rm M }   [   \cdot ~ (  |y\>\<y| \otimes  I_{\rm S}   )] \Big)  \, .
	 \end{align}

	The operations performed by machines with purely classical memory will be called  \emph{measure-and-operate (MO) strategies}.    Combined together, the ``measure'' channel $\map E_\theta$ and the ``operate'' channel $\map R_\theta$ give a single quantum channel $\map C_{\theta, \rm MO}$, of the form 
    \begin{align}\label{CMO}
		{ \map C}_{\theta}   (\rho)   =  \sum_{y\in \set Y}  \,    \map O_{\theta,y}  \Big(  \Tr_{\overline {\rm S}} \big[(P_{\theta, y}\otimes I_{\rm S})  \,  \rho \big]  \Big)   \, ,  
    \end{align} 
    where $\Tr_{\overline {\rm S}}$ denotes the partial trace over all systems except system $\rm S$.

	In the following, we will  solve the optimisations in Problems \ref{prob:channelonly} and \ref{prob:optimiseall} under the constraint that the channel $\map C_\theta$ is of the  $\rm MO$ form  (\ref{CMO}).     By definition, the optimal {\rm MO} fidelities are by definition no larger than the optimal quantum fidelities derived in the previous Section. 
	% since the latter correspond to an optimisation over  {\em all} quantum channels $\map C_\theta$. 
	
	\subsection{Structure of the optimal MO strategy for Problem \ref{prob:optimiseall}} 
   	The structure   of the optimal MO strategy for Problem \ref{prob:optimiseall} is summarized by the following Theorem, proven in Appendix \ref{app:optMOstructure}.

	\begin{theo}\label{theo:optimalMOstrategy}
	The optimal  MO strategy for  learning the gate $V_{\theta,g}  =  U_g V_\theta  U_g^\dag$ from the  gate $U_g^{(j)}$ has the following features: 
	\begin{enumerate}
	\item no auxiliary system is needed
	\item the optimal probe state is an eigenstate of $J_z$, denoted as $|j,m_\theta\>$
	\item  the outcome of the optimal POVM is an element of the rotation group $\grp{SO} (3)$, denoted as $\hat{g}$
	\item  the optimal POVM $(P_{\theta,  g})_{g\in\grp {SO}(3)}$ is  rotationally covariant  \cite{holevo2011probabilistic}, and has the form 
	\begin{align}\label{covPOVM}
		P_{\theta, \hat{g}}   =  (2j+1) ~    U^{(j)}_{\hat{g}}   |\xi_\theta\>\<\xi_\theta|  \, U_{\hat{g}}^{(j) \, \dag}      \, , 
	\end{align} 
	where $|\xi_\theta\> $ is a unit vector 
		\item the optimal conditional operation   has the form $\map O_{\theta,  {\hat{g}}}  =  \map U_{\hat{g}}^{(j) } \circ \map O_\theta  \circ \map U_{\hat{g}}^{(j) \dag}$, where $\map O_\theta$ is a fixed channel acting on the target qubit.    

	\end{enumerate}
	\end{theo}

In the following we will maximise the gate fidelity over all MO strategies with the features described by Theorem \ref{theo:optimalMOstrategy}. 	For convenience we will  express  the gate fidelity in terms of   the entanglement fidelity [cf. Equation  (\ref{horodecki})].

	\subsection{Choi operator formulation}

  For an optimal strategy as in Theorem \ref{theo:optimalMOstrategy},  the entanglement fidelity takes the form
    \begin{align}\label{insomma}
    \nonumber F^{\rm (e)}_{\rm 2,MO}(j,\theta) &= (2j+1) \, \int \d g \, \left|\<\xi_\theta|U_{\hat{g}}^{ (j)\dagger}   U_g^{(j)}|j,m_\theta\> \right|^2 ~  \frac{\<   \Phi_{\theta,g}^+  |  O_{\theta,  {\hat{g}}} |\Phi^+_{\theta,g}\> \ }2   \\
    &= (2j+1)  \,  \int  \d g  \,     \left| \<  \xi_\theta  |  U_g^{(j)}  |j,m_\theta\> \right|^2 ~  \frac{\<   \Phi_{\theta,g}^+  |  O_{\theta} |\Phi^+_{\theta,g}\> \ }2 
    \, , 
    \end{align}
	where $O_{\theta,  {\hat{g}}}$ is the Choi operator of the channel $\map O_{\theta, {\hat{g}}}$, $O_{\theta}$ is the Choi operator of the channel $\map O_{\theta}$, and $|\Phi^+_{\theta,g}\>  :  =   (V_{\theta,g} \otimes I_{\rm R}) \, |\Phi^+\>$.

	Our goal is to maximise the entanglement fidelity (\ref{insomma}) over all values of $m_\theta$, over all unit vectors $|\xi_\theta\>$, and over all Choi operators $O_\theta$.   To this purpose, the key observation is that the Choi operator $O_\theta$ can be chosen to be real in a suitable basis. 
	Specifically, we have the following 
    \begin{prop}\label{prop:real}
		The Choi operator $O_\theta$  maximizing the fidelity (\ref{insomma}) can be chosen to be real in the Bell basis 
    \begin{align}
	{\sf B}_{\rm Bell}  =\Big \{    |\Phi^+\>  \, ,  \,  i (\sigma_x \otimes I)  |\Phi^+\> \, ,  \,  i (\sigma_y \otimes I)  |\Phi^+\> \, ,\,  i (\sigma_z \otimes I)  |\Phi^+\>  \Big\}  
	\end{align}
 
	\end{prop}
	\Proof    Every unitary $V_{\theta,g}  =    U_g V_\theta  U_g$ is a real linear combination of the matrices $ I,  i\sigma_x,  i\sigma_y,   $ and $i\sigma_z$.  Hence, every vector $|\Phi^+_{\theta,g}\>  =  (V_{\theta,g} \otimes I)  |\Phi^+\>$ is a real linear combination of the vectors    $|\Phi^+\>$,  $i|\Psi^+\>   =   ( i\sigma_x\otimes I)  |\Phi^+\>$, $|\Psi^-\>   =   ( i\sigma_y\otimes I)  |\Phi^+\>$, and $i|\Phi^-\>   =   ( i\sigma_z\otimes I)  |\Phi^+\>$.   Since the fidelity depends on the Choi operator $O_\theta$ only through the matrix elements $\<\Phi^+_{\theta,g}|  O_\theta|\Phi^+_{\theta,g}\>$, the optimal Choi operator can be chosen to be real in the same basis as the vectors $|\Phi^+_{\theta,g}\>$. \qed

	Thanks to Proposition \ref{prop:real}, the maximization of the fidelity can be restricted to the set of Choi operators that are real in the Bell basis.      This set of Choi operators can be equivalently characterized as the set of Choi operators of unital channels,  {\em i.e.} quantum channels mapping the identity operator to itself.   Indeed, we have the following  
	\begin{prop}
	A qubit channel is unital if and only if its Choi operator    is  real in the Bell basis  
	\begin{align}
		{\sf B}_{\rm Bell}=  \Big\{   |\Phi_0\>  =  |\Phi^+  \>  \,  , \,     |\Phi_1\>  =    i (  \sigma_x \otimes  I) |\Phi^+  \>   \, ,  \,    |\Phi_2\>  =    i (   \sigma_y\otimes I) |\Phi^+  \>   \, ,\,    |\Phi_3\>  =    i (  \sigma_z\otimes I) |\Phi^+  \>    \Big\}\, . 
	\end{align} 
	\end{prop}

	\Proof    If a qubit channel is unital, then it is a convex combination of unitary channels \cite{landau1993birkhoff}.  For every unitary channel, the corresponding Choi operator   is real in the Bell basis. Indeed, every unitary channel   has a Kraus decomposition with a single unitary operator  of the form $U  =   \cos \frac \tau 2  \,  I  -   i  \sin \frac \tau 2\,  \st n  \cdot \bs \sigma$, with $\tau  \in  [0,2\pi)$ and  $\st n   \in \R^3$. Hence, the Choi operator $ 2 \, (U\otimes I)   |\Phi^+\>\<\Phi^+ |    (U\otimes I)^\dag $ 
	is real in the Bell basis. Since the set of real Choi operators is convex, every unital channel is contained in it. 

	Conversely,   suppose that a channel $\map C$ has a Choi operator $C$  that is real in the Bell basis, {\em i.e.}  $C  =  \sum_{k,l}   \, C_{kl} \,  |\Phi_k\>\<\Phi_l|$, for some real symmetric matrix  $(C_{kl})$.    Then, one has 
	\begin{align}
		\nonumber  \map C (I)   &   =  \Tr_{\rm in}[  C ]   \\
		\nonumber  &  =   \, C_{00} \,    \frac I 2    +   \sum_{1\le k\le 3}   \,  C_{0k}   \,      \left ( \frac{  -i   \sigma_k    +   i \sigma_k }2   \right)   +   \sum_{1\le k\le   l  \le 3}  \, C_{kl}  \,  \left (\frac{  \sigma_k \sigma_l + \sigma_l\sigma_k }4 \right) \\
		\nonumber  &  =   \frac {C_{00}}2 \,   I   +  \sum_{1\le k\le   l  \le 3}  \, C_{kl}  \,  \frac{ \delta_{kl}  }2 \, I \\
		\nonumber &  =  \frac{\sum_{i=0}^3    C_{ii}}2  \, I \\
		&  =  I    \, ,
	\end{align}
	the last equality following from the relation  $2   =  \Tr[  I]   =  \Tr[\map C(I)]   =  \Tr [  C]   =  \sum_{i=0}^3 C_{ii}$.  Hence, the channel $\map C$ is unital. \qed

	Since the fidelity is a linear function, its maximization can be restricted to the extreme points of the set of unital  channels. 
	For qubits, such extreme points are  unitary channels \cite{landau1993birkhoff}. Hence,  we obtained the following  

	\begin{theo}\label{theo:optimalMOunitary}
	    The quantum channel  $\map O_\theta$  maximizing the fidelity (\ref{insomma}) can be chosen to be unitary without loss of generality. 
	\end{theo} 

	Thanks to Theorem \ref{theo:optimalMOunitary}, the optimal entanglement fidelity (\ref{insomma}) can be expressed as 
    \begin{align}\label{insomma2}
	    F^{\rm (e)}_{\rm 2,MO,opt}(j,\theta)&=  \max_{m_\theta  \in   \{-j, \dots, j\}}  ~\max_{|\xi_\theta\>:  \| |\xi_\theta\>\|=1} ~ \max_{  W_\theta:   W_\theta^\dag W_\theta = I}    \Big\{ (2j+1)  \,  \int  \d g  \,     \left| \<  \xi_\theta  |  U^{(j)}_g |j,m_\theta\> \right|^2 ~    \big|  \< \Phi^+_{W_\theta}  |     \Phi_{\theta,g}^+    \> \big|^2  \Big\}  \, , 
	\end{align}
	where $W_\theta$ is a suitable unitary and $|\Phi^+_{W_\theta}  \>   :=  ( W_\theta  \otimes I_{\rm R})  \,  |\Phi^+\>$.

	\iffalse More compactly, we can also write 
	\begin{align}\label{2am}
		F^{\rm (e)}_{\rm MO}(j,\theta)=  \max_{m  \in   \{-j, \dots, j\}}  ~\max_{|\xi_\theta\>:  \| |\xi_\theta\>\|=1} ~ \max_{  W_\theta:   W_\theta^\dag W_\theta = I}    ~   \Big (   \<   \xi_\theta  |  \otimes \<  \Phi^+_{W_\theta}|  \Big)   \,   \Omega_m\,     \Big (   |   \xi_\theta  \>  \otimes | \Phi^+_{W_\theta}\>  \Big)  \, ,
	\end{align}
	with 
	\begin{align}
		\Omega_m =  (2j+1)  \,   \int \d g \,  \Big( \map U_g^{(j)}\otimes \map U_g \otimes  \overline  {\map  U}_g \Big)\,  \Big(   |j,m\>\<j,m|  \otimes  |\Phi_{\theta}^+\>\<\Phi^+_\theta|   \Big) \, .
	\end{align} 
\fi 
	The optimization can be further simplified using the following observation: 
	\begin{prop} 
	The unitary gate $W_\theta$ maximizing the fidelity (\ref{insomma}) can be chosen without loss of generality to be a rotation about the $z$ axis.  
	\end{prop}
	\Proof   Every unitary $W_\theta$ can be written as $  W_\theta  =   U_h  V_{\theta^{'}}  U_h^\dag$, where $V_{\theta^{'}}$ is a rotation about the $z$ axis by an angle $\theta^{'}$, and $h$ is the rotation that transforms the $z$ axis into the rotation axis of $W_\theta$.
	Hence,  the corresponding state can be written as $|\Phi^+_{W_\theta} \>   =   (U_h\otimes \overline U_h) \,  |\Phi^+_{V_{\theta^{'}}}\>$.
	%\Proof   Every unitary $W_\theta$ can be written as $  W_\theta  =   U_h  W_\theta'  U_h^\dag$, where $W_\theta'$ is a rotation about the $z$ axis and $h$ is the rotation that transforms the $z$ axis into the rotation axis of $W_\theta$.    Hence,  the corresponding state can be written as $|\Phi^+_{W_\theta} \>   =   (U_h\otimes \overline U_h) \,  |\Phi^+_{W_\theta'}\>$.  

\iffalse
Then, we have the relation 
\begin{align}
\nonumber   \Big (   \<   \xi_\theta  |  \otimes \<  \Phi^+_{W_\theta}|  \Big)   \,   \Omega_m\,     \Big (   |   \xi_\theta  \>  \otimes | \Phi^+_{W_\theta}\>  \Big)    &=  \Big (   \<   \xi_\theta  |  \otimes \<  \Phi^+_{W_\theta'}|  \Big)   \,   \Big (\map I_j  \otimes \map U_h^\dag \otimes \map U_h^T \Big)  \Omega_m\,     \Big (   |   \xi_\theta  \>  \otimes | \Phi^+_{W_\theta'}\>  \Big)   \\
&=  \Big (   \<   \xi_\theta  |  \otimes \<  \Phi^+_{W_\theta'}|  \Big)   \,   \Big (\map U_j  \otimes \map I\otimes \map I \Big)  \Omega_m\,     \Big (   |   \xi_\theta  \>  \otimes | \Phi^+_{W_\theta'}\>  \Big)  
\end{align}
\fi

	Using this fact, the optimal MO fidelity can be rewritten as  
    \begin{align}\label{insomma3}
	    \nonumber  F^{\rm (e)}_{\rm 2,MO,opt}(j,\theta)&=   \max_{m_\theta  \in   \{-j, \dots, j\}}  ~\max_{|\xi_\theta\>:  \| |\xi_\theta\>\|=1} ~ \max_{  W_\theta:   W_\theta^\dag W_\theta = I}    
	    \Big\{ (2j+1)  \,  \int  \d g  \,        \left| \<  \xi_\theta  |  U_g^{(j)} | j,m_\theta\> \right|^2 ~    \big|  \< \Phi^+_{W_\theta}  |   \Phi_{\theta,g}^+    \> \big|^2  \Big\} 	       \\
		\nonumber 	    &  = \max_{m  \in   \{-j, \dots, j\}}  ~\max_{|\xi_\theta\>:  \| |\xi_\theta\>\|=1} ~ \max_{  W_\theta:   W_\theta^\dag W_\theta = I}    
	    \Big\{ (2j+1)  \,  \int  \d g  \,     \left| \<  \xi_\theta| U_h^{ (j)\dagger}  U_g^{(j)}  |j,m_\theta\> \right|^2 ~    \big|  \<   \Phi^+_{V_{\theta^{'}}}  | \,  (U_h^\dagger\otimes U_h^{T}) \, |\Phi_{\theta,g}^+\> \big|^2  \Big\}  \\
	    \nonumber  &=  \max_{m_\theta  \in   \{-j, \dots, j\}}  ~\max_{|\xi_\theta\>:  \| |\xi_\theta\>\|=1} ~ \max_{h\in \grp{SO}(3)}~  \max_{  {V_{\theta^{'}}}:   	V_{\theta^{'}}^\dagger V_{\theta^{'}} = I}   
		\Big\{ (2j+1)  \,  \int  \d g  \,     | \<  \xi_\theta  |  U_{h^{-1} g }^{(j)}  |j,m_\theta\> |^2 ~    \big|  \<   \Phi^+_{V_{\theta^{'}}}   |\Phi_{\theta,h^{-1}g}^+\> \big|^2  \Big\}  \\
		&  =  \max_{m_\theta  \in   \{-j, \dots, j\}}  ~\max_{|\xi_\theta\>:  \| |\xi_\theta\>\|=1} ~  \max_{  {V_{\theta^{'}}}:   	V_{\theta^{'}}^\dagger V_{\theta^{'}} = I}
		\Big\{ (2j+1)  \,  \int  \d g'  \,    \left | \<  \xi_\theta   |  U_{g'}^{(j)}  |j,m_\theta\> \right|^2 ~    \big|  \<   \Phi^+_{V_{\theta^{'}}}   |\Phi_{\theta,g'}^+\> \big|^2  \Big\} \, ,
	    %&  =  \max_{m  \in   \{-j, \dots, j\}}  ~\max_{|\xi_\theta'\>:  \| |\xi_\theta'\>\|=1} ~ \max_{  {V_{\theta^{'}}}:   	V_{\theta^{'}}^\dagger V_{\theta^{'}} = I}
	    %\Big\{ (2j+1)  \,  \int  \d g'  \,     | \<  \xi_\theta'      |j,m\>_{g'} |^2 ~    \big|  \<   \Phi^+_{V_{\theta^{'}}}   |\Phi_{g'}^+\> \big|^2  \Big\}   \, .   
	\end{align}  
	(here $U^{T}_h$ denotes the transpose of the matrix $U_h$.) 
    The last equation shows that the maximisation of the fidelity  can be reduced to rotations about the $z$ axis. \qed

	At this point, it  remains to  maximise the  fidelity (\ref{insomma3}) over $m_\theta$,  $\xi_\theta$, and $V_{\theta^{'}}$.    The result of the  optimization is summarised in the following, while the details are provided in Appendix \ref{app:MOoptimization}.

   	\subsection{Optimal MO strategy for $j\not  = 1$}
   	For $j\neq1$,  it turns out that the quantum benchmarks for Problems \ref{prob:channelonly} and \ref{prob:optimiseall} coincide.

	\begin{theo}\label{theo:5}
		For  $j\not = 1$, the optimal  probe  state   for learning the gate $V_{\theta,g} = U_g V_\theta  U_g^\dag$ from the gate $U_g^{(j)}$ by MO operations is $|j,j\>$  for every value of $\theta$.   For both Problems \ref{prob:channelonly} and \ref{prob:optimiseall}, 
		%optimal entanglement fidelity is 
		%		\begin{align}\label{FidOptimal}
		%		F^{(\rm e)}_{\rm opt}(j ,\theta)=    \frac{1 + \sqrt{ 1  +    \frac{2j+1}{j^2}  \, \left (\cos \frac \theta 2 \right)^2  }   +   \frac{2j+1}{2j^2}  \, \left (\cos \frac \theta 2 \right)^2   }{2 \, (1+ \frac 1{2j})^2}  \, .
		%	\end{align}  
		%	Equivalently, the 
		optimal MO   fidelity  is
	  \begin{align}\label{MO_Fopt}
		F_{\rm  MO,opt}&(j,\theta)   = \dfrac{4j+4+(2j+1)\cos(\theta-\theta^{'})}{3(2j+3)}  +\dfrac{(2j+1)(\cos\theta+\cos\theta^{'})+\cos(\theta+\theta^{'})+1}{3(j+1)(2j+3)}  \, .
	\end{align}	
		 and has the asymptotic expression 
	 \begin{align}\label{FMOapp}
        F_{\rm MO,opt} (j,\theta) = 1-\dfrac{  2    \left(1 - \cos\theta \right) }{3j} 
        %\dfrac{\cos^{2}\theta -4\cos\theta+3}{3j^{2}}
        +O\left(\dfrac{1}{j^2}\right) \, .
    \end{align}	
    The optimal MO strategy  consists in 
     \begin{enumerate}
    \item  measuring the probe with the POVM  $P_{\hat g}  =  (2j+1)~     U_{\hat g}^{(j)}    |j,j\>\<j,j|   U_{\hat g}^{(j) \dag}$, and
    \item  rotating the target qubit  about  the rotated $z$-axis  $\hat g \, \st e_z$       
    by the angle 
    \begin{align}\label{tauinapp}
		\theta^{'} = \arccot  \left[  \cot \theta    +    \dfrac{2  \cos\theta+2j+1}{(2j^{2}+3j)\sin\theta}\right]  +  s(\theta)
    \end{align}
    where $s(\theta)  =  0$ for $\theta \in  [0,\pi]$, and $s(\theta)  =  \pi$ for $\theta  \in  (\pi,2\pi)$. 
    \end{enumerate}
			\end{theo}

        Note that the probe state and the measurement are both independent of the rotation angle $\theta$. This means that the machine can be trained optimally even before the value of the rotation angle has been decided. 
          The operations in the training phase coincide with the optimal estimation  strategy for directions, derived in the classic work by Holevo \cite{holevo2011probabilistic}.

        %The reason for this robustness is the symmetry of the problem, which imposes a strong constraint on the structure of the optimal states and measurements.   For sufficiently large $j$, these constraints implies that one estimation strategy is optimal for multiple tasks. 
            	
The optimal MO strategy can be implemented by a learning machine with a purely classical memory.   The size of the classical memory can be chosen without loss of generality to be $\lceil 2 \log  (2j+1)  \rceil$ bits.  This is because the fidelity is a linear function of the POVM, and therefore its maximum is attained by an extreme point of the convex set of all  POVMs with outcomes in $\grp {SO} (3)$.  The extreme points of such set consist of POVMs that assign non-zero probability to at most $(2j+1)^2$ rotations  \cite{chiribella2007how}.  Hence, the optimal POVM in Theorem \ref{theo:5} can be replaced  by another, equally optimal POVM with at most $(2j+1)^2$ outcomes, which can be stored into a classical memory of $\lceil 2 \log  (2j+1)  \rceil$ bits.

 \begin{figure}[ht]
    	    \centering
    	    \includegraphics[width=0.52\textwidth]{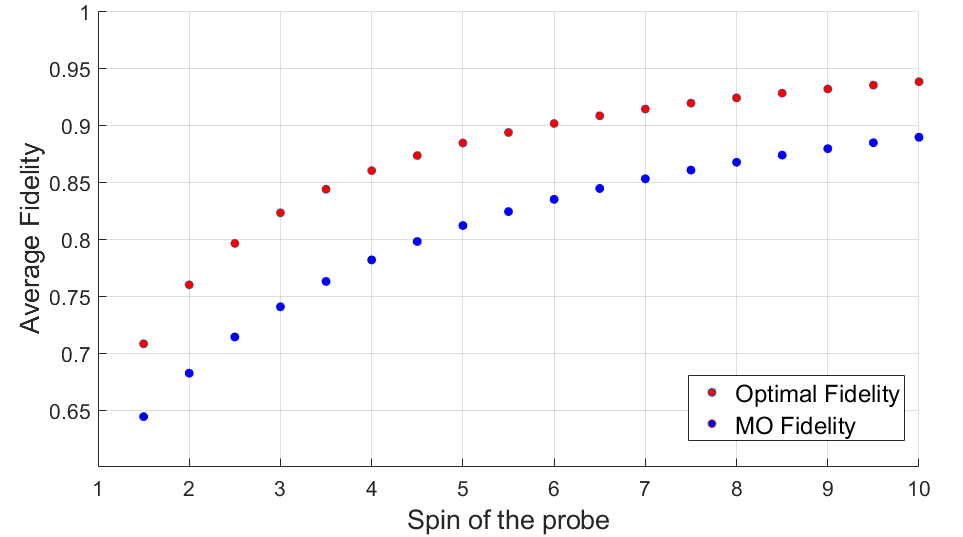}
    	    \caption{\footnotesize
    		    \textbf{Benchmark for quantum learning of rotation gates.}   The quantum benchmark (blue dots) and the optimal quantum fidelity (red dots) are plotted for rotations of $180$ degrees in a function of the spin size, with $j$ ranging from $3/2$ to $10$.}
    	    \label{fig:benchmark}
    \end{figure}
    
	A plot of  the  MO fidelity  and of  the optimal quantum fidelity  is provided in Figure \ref{fig:benchmark}.  
    % and for  $\theta  =  \pi$. 
	Note that  the error (one minus fidelity)    goes to zero in both cases,  but the rate for quantum strategies is twice as fast, as one can see by comparing Equations  (\ref{Foptapp}) and (\ref{FMOapp}).

	\subsection{Optimal MO strategies for $j = 1$}\label{subsect:MOj=1}

The $j = 1$ case exhibits an anomalous behaviour when the rotation angle approaches $\pi$.  For   $|\pi - \theta| > 0.303\pi$,   Problems \ref{prob:channelonly} and \ref{prob:optimiseall}  have the same optimal MO fidelity, and the same  optimal MO strategy,  described in Theorem \ref{theo:5}. 
For  $|\pi - \theta| \leqslant 0.303\pi$, the optimal  the optimal MO fidelities become different. 
For $|\theta-\pi|\le \delta_{1}$,  the optimal average fidelity for  Problem \ref{prob:channelonly} is 
\begin{align}\label{FMOnew1}
			F_{1, \rm   MO, opt}(j =1,\theta)=  \dfrac{8+3\cos(\theta-\theta^{'})}{15}  +\dfrac{3(\cos\theta+\cos\theta^{'})+\cos(\theta+\theta^{'})+1}{30}    \, ,
		\end{align} 
		with
			\begin{align}
\theta^{'} = \arccot  \left[  \cot \theta    +    \dfrac{2  \cos\theta+3}{5\sin\theta}\right]  +  s(\theta)\, .
    \end{align}
    corresponding to  Equation (\ref{MO_Fopt})  with $j=1$.
		 The MO strategy is still the one described in Theorem \ref{theo:5}.

For Problem \ref{prob:optimiseall}, the optimal probe states  transitions  from $|1,1\>$ to $|1,0\>$, and the optimal fidelity becomes 
	\begin{align}\label{Fid1avebis}
		F_{\rm 2, MO, opt}(j=1,\theta)     =  \frac 13  + \frac 25 \,  \left(\sin \frac \theta 2\right)^2   \, .
	\end{align}

The optimal MO strategy consists of \begin{enumerate} 
    \item  Measuring the memory with the POVM operators  $P_{\hat g}  =  (2j+1) ~   U_{\hat g}^{(j)}   |1,0\>\<1,0|  U_{\hat g}^{(j) \dag}$
    \item  Rotating the target qubit about the axis $\st n  =  g \, \st e_z$  by an angle   $\pi$, independently of $\theta$. 
    \end{enumerate}
    Physically, the optimal  POVM  can be interpreted as a randomisation of the projective measurement that projects the spin-$1$ particle along the three Cartesian axes $x,y$, and $z$ \cite{chiribella2007how}.  This projective measurement corresponds to the orthonormal basis $\{  |x\>, |y\>  ,|z\>\}$ for $\C^3$ defined by $|z\>  : =  |1,0\>$, $|x\>  :  =  (|1,1\>  +  |1,-1\>)/\sqrt 2$,  and $|y\>  :  =  (|1,1\>  -  |1,-1\>)/\sqrt 2$.  In the language of atomic physics, $|x\>$, $|y\>$, and $|z\>$ are the $p$-orbitals aligned in the directions $x,y$, and $z$, respectively.  Since the fidelity is a linear function of the POVM, the optimal POVM  $P_{\hat g}  =  (2j+1)\,    U_{\hat g}^{(j)}   |1,0\>\<1,0|  U_{\hat g}^{(j)  \dag}$ can be replaced by an equally optimal POVM  based on the projective measurement of $\{  |x\>, |y\>  ,|z\>\}$, followed by a rotation by $\pi$ about the Cartesian axis identified by the measurement outcome.  In this discretised version of the MO strategy, the learning machine only needs a classical memory of $2$ bits.

	\section{Persistence of the quantum advantage}\label{sec:longevity} 

	We have seen that a machine equipped with a quantum memory can outperform every classical machine at the task of learning rotations about an unknown axis.    Still, our analysis was restricted to the scenario where the quantum process accesses its memory only once, with the goal of reproducing a single use of the target gate.  In the following we will study how the performance depends  on the number of required  executions of the target gate.

	Let us focus on the regular case $j >1$, where the optimal strategies for Problems \ref{prob:channelonly} and \ref{prob:optimiseall} coincide, and the channel is realised by setting  up a Heisenberg interaction between the memory and the target qubit.   %Physically, the  Heisenberg interaction transfers  information  from the memory to the  target qubit, which undergoes an effective evolution determined by the memory state.     This leads to  a backreaction effect, whereby the memory gradually deteriorates: every time the memory is accessed, the machine ``forgets" some amount of the information that was previously learnt.    This kind of {\em quantum forgetting} is in stark contrast with the everyday experience of learning in the classical world, where executing a task reinforces the memory.     In the context of quantum reference frames, the degradation of the reference frame information due to the interaction with a target system was studied in the task of measuring a quantum particle along the direction indicated by another particle  \cite{bartlett2006degradation,bartlett2007degradation,poulin2007dynamics}. 
	%
	%After many accesses,  the information content of the memory will be depleted.   
	An important question is how many times the memory can  be accessed before the accuracy drops below a certain threshold.  In the  context of quantum reference frames, the maximum number of accesses such that the fidelity is above  threshold  was called the  \emph{longevity}  in Ref. \cite{bartlett2006degradation}.  Another important question is how many times the memory can be accessed before the  quantum advantage is lost.   
	%Technically, the question is to find the maximum number $n$ such that the fidelity for the execution of the target gate is larger than the MO fidelity. 
		The  maximum number of accesses for which the fidelity is above the quantum benchmark (\ref{MO_Fopt}) will be called  {\em persistence of the quantum advantage} in the following. 

	%, borrowing the terminology of Ref.   \cite{bartlett2006degradation}.  

	Suppose that the joint evolution of memory and target is described by the same unitary gate  at every step. Assuming the gate to be of the form  of Eq.  (\ref{optimalU}) for  some fixed function $f(\theta)$,  we obtain the   close-form  expression 
    \begin{align}\label{close}
        F(j,\theta,n) = 1 - \dfrac{1-\cos\theta}{3j}\cdot\dfrac{n(1-\cos\theta)+j}{j} + O\left(\dfrac{1}{j^2}\right)
    \end{align}
    quantifying the average fidelity at the leading order in $j$ (see Appendix \ref{app:longevity} for the derivation). 
    %Note that the error grows linearly with the number of recycling steps. 
    From this expression one can see that the longevity grows as $j^2$.  
    %A quadratic scaling was also observed in Ref. \cite{bartlett2006degradation}, for the different problem of controlling the direction of a polarization measurement.         
    However, the persistence of the quantum advantage is much shorter: comparing the fidelity (\ref{close}) with the MO fidelity (\ref{FMOapp}), we  find that the quantum advantage disappears when the number of repetitions is larger than 
    \begin{align}\label{long}
        N  (j,\theta)=  \dfrac{j}{1-\cos\theta} + O\left(1\right)    \, .
    \end{align}
	One could  also consider  more elaborate strategies  where the interaction time between memory and target  is optimised at every step.    However, we find that these strategies do not increase the longevity nor the persistence of the quantum advantage in the large $j$ limit. 
	%, which remains equal to  $  L(j,\theta)  =j/{(1-\cos\theta)} + O(1)$.
    
    %For approximate proof, we show an example of $j=10^6$. Using the fact that the distribution of $U_{\bf n}^{(j)}|j,m\>\<j,m|U_{\bf n}^{(j)\dagger}$ is nearly constant, which is nearly a geometric progression, after reusing for $\lceil(j-1)/2\rceil$ times, and with probablility larger than $99.99\%$ distributed in first 10 terms.
    
    %Now using just first 20 terms to form the transfer matrix $T^{(j)}$ with elements:
    %    \begin{align*}
    %    	\begin{cases}
    %    		c_{m-1,m}=\dfrac{1}{(1+2j)^{2}}\left[(j-m)(1+j+m)(1-\cos\varphi)\right]\\
    %    		c_{m,m}=\dfrac{1}{(1+2j)^{2}}\left[2j^{2}+2m^{2}+2j+1+2\cos\varphi(j^{2}+j-m^2)\right]\\
    %    		c_{m+1,m}=\dfrac{1}{(1+2j)^{2}}\left[(j+m)(1+j-m)(1-\cos\varphi)\right]
    %    	\end{cases}
    %    \end{align*}
    %to calculate $p(n-1,m)$. Finally we will get $F(j,\theta,n)_{\rm opt}  >  F_{\rm MO}(j,\theta)_{\rm opt}$ when $n \leqslant 5\times10^{5}-293$.

\medskip

	\section{Robustness to thermal noise}\label{sec:robustness}
	In Problem \ref{prob:channelonly}, we made the simplifying  assumption that  the unknown direction $\st n$ is imprinted into  the pure spin-coherent state $|j,j\>_{\st n}$,  regarded as the low-temperature approximation of the thermal state of the magnetic dipole Hamiltonian.    An interesting question is how this approximation affects our discussion of the quantum advantage.   In the following we will address this question in the large $j$ limit, showing that quantum memories are useful  whenever the magnetic energy is sufficiently large compared to the thermal fluctuations.  	
	
The thermal states of the Hamiltonian  	$  H  =  -\mu  \, \st B  \cdot  \st J$ can be written as 
	\begin{align}\label{mixed}
	\rho_{\gamma,\st n}    =  \frac{ \sinh \gamma  }{\sinh [(2j+1) \gamma]}   \sum_m   e^{     2 \, \gamma  \, m}  |j,m\>\<j,m|_{\bf n}  \, , \qquad \gamma  =  \frac{\mu |\st B|}{2  k_{\rm B}  T} \, ,
	\end{align} 
where  $T$ is the temperature and  $k_{\rm B}$ is  the Boltzmann constant. The spin coherent state $|j,j\>_{\st n}$ is  retrieved in the low temperature  ($\gamma \to \infty$) limit, as  one has $\lim_{\gamma\to \infty} \rho_{\gamma,\st n}   =  |j,j\>_{\st n} \<j,j|_{\st n}$.  

Now, suppose that the  learning strategy designed for  the spin coherent state   $|j,j\>_{\st n}$  is adopted for the  mixed state  $\rho_{\gamma,\st n}$.   In  Appendix  \ref{app:robustness}, we show that the average fidelity has the asymptotic expression 
	\begin{align}\label{robustclose}
		F_1(j,\theta,\gamma) = 1 - \dfrac{1-\cos\theta}{3j  \tanh \gamma}  + O\left(\dfrac{1}{j^2}\right) \, . 
	\end{align}

The above fidelity can be compared the benchmark in  Equation (\ref{FMOapp}), which quantifies the maximum fidelity achievable with classical memories. Note that  Equation (\ref{FMOapp}) provides the benchmark for both Problems \ref{prob:channelonly} and \ref{prob:optimiseall}, meaning that the benchmark applies to every  pure probe state of the form $|\psi_{\theta,g}\>  =  U_g^{(g)} |\psi\>$, and by convexity, to every mixed probe state of the form $\rho_g  =  U_g^{(j)}  \rho  U_g^{(j) \dag}$.  In particular, it applies to the thermal states $\rho_{\gamma,\st n}$, as the average fidelity over all directions $\st n$ is equal to the average fidelity over all rotations $g$.  
Comparing the fidelity (\ref{robustclose}) with the benchmark in  Equation (\ref{FMOapp}), we obtain that the quantum strategy outperforms all classical strategies whenever $\tanh \gamma$ is larger than $1/2$, corresponding to the condition $\gamma >  \frac 12 \ln  3  \approx 0.55 $.   Hence, the quantum  advantage persists whenever the magnetic energy $\mu  |\st B|$ is larger than $1.1$ times the thermal energy $k_{\rm B}  T$.

Note that the quantum benchmark   in  Equation (\ref{FMOapp}) is the optimal fidelity achievable with arbitrary probe states. If one further enforces the condition that the the probe state be thermal, then the value of the benchmark would be even lower, thereby extending the set of temperatures for which the quantum memory offers an advantage. 

Note also that the above discussion applies to a variant of Problem \ref{prob:optimiseall} where the probe is subject to thermal noise before the action of the training gate $U_g^{(j)}$, resulting into a mixed input state $\rho_\gamma :  =  \rho_{\gamma,\st e_z}$.  Also in this setting, the quantum memory offers a provable advantage when the parameter $\gamma$  is larger than $\frac 12 \ln  3$.

	\section{Learning higher dimensional gates}\label{sec:larger}   
	Our result establishes the existence of a quantum advantage  for  learning single-qubit rotations about an unknown axis.  This finding  is conceptually important, because the advantage for single qubits implies an advantage  of coherent learning for quantum systems of arbitrary  dimension. Indeed, one can  immediately prove the advantage by using the qubit benchmark for gates that act nontrivially only in a fixed two-dimensional subspace.
        
	Our results also  give a heuristic for the problem of learning rotation gates on  higher dimensional spins.   
    The idea is to encode the rotation axis in a  spin coherent state  and to let  the memory and target spin interact as closed system. Explicitly, we make two spin systems undergo the Heisenberg interaction $U_\theta^{(k)}  =     \exp  \left[    -i \theta \,   \,   2\st J \cdot \st K/ (2j+1)  \right]$,   %Specifically, we choose the gate 
    %  \begin{align}     $U_{\theta}=\bigoplus_{l=j-k}^{j+k}  \,  e^{-i\theta l}  \,  P_{l}$,
    %  \textcolor{blue}{Maybe the fidelity improves if we choose  $U_{\theta}=\bigoplus_{l=j-k}^{j+k}  \,  e^{-i\theta l (l+1)}  \,  P_{l}$. Can you check what is the fidelity in this case? If it is, can you replace this part with the improved results?} 
    %\end{align} 
    where $\st K  =  (K_x, K_y,  K_z)$ are the spin operators of the target spin.  %where $k$ is the spin of the target and $P_l$ is the  projector on the subspace with quantum number $l$ for the total angular momentum.  
    %The  above  gate   is  the direct generalization to higher spins of the unitary gate in Eq.(\ref{optimalU}).  
    Using the unitary gate $U_\theta^{(k)}$, in Appendix \ref{app:spink} we obtain the average fidelity 
    \begin{align}
       	F_{\rm Hei}(j,k,\theta)  =   1-\dfrac{k(2k+1) (1-\cos\theta)}{3j}    \, ,         
    \end{align}
    in the large $j$ limit.  Remarkably, the error grows \emph{quadratically}---rather than linearly---with the size of the target spin: in order to ensure high fidelity, the size of the memory  must be large compared to the square of the size of the target system.  The same conclusion holds for the worst-case fidelity, which has the asymptotic  expression  
    \begin{align}
        F_{\rm w,Hei}(j,k,\theta)  =  1  - \frac  { [k(k+1)  +  c (k)]\,  (1-\cos\theta)}{j}    \,, 
    \end{align} 
    with $c(k)  = 0 $ for even $k$ and $c(k)=  1/4$ for odd $k$.

	The quantum strategy exhibits an advantage over the MO strategy consisting in measuring the direction $\st n$ from the spin coherent state pointing in direction $\st n$  and performing a rotation based on the outcome. 
    Again, we find that the error of the quantum strategy vanishes in the macroscopic limit of large memory systems, at a rate twice as fast  than  the error of the classical strategy (see Appendix \ref{app:spink} for more details).    It is an open question whether the above quantum and MO strategies are optimal for arbitrary $k>1/2$.
     
    %  It is tempting to conjecture that this is a general rule that applies to the problem of controlling unitary gates in the macroscopic limit. 

\medskip

	\section{Conclusions}\label{sec:conclusions}    
	We determined the ultimate accuracy for the task of learning a rotation of a desired angle $\theta$  about an unknown axis, imprinted in the state of a spin-$j$ particle. 
	  In this task, we found that quantum memories enhance the learning performance for every $j>1$ and for every rotation angle $\theta\not = 0$. Specifically, we found that a quantum machine with a memory of $\lceil \log( 2 j+1)  \rceil$ qubits outperforms all learning machines with classical memory of arbitrarily large size. 
	  
We found that the advantage of the quantum memory persists even when the memory is accessed multiple times, as long as the total number of accesses is at most linear in the spin size.    
	Quite interestingly, we observe a relation between the persistence and the size of the advantage:  in the large $j$ limit, the quantum advantage is of size $O(1/j)$ and persists when the memory is accessed  for $O(j)$ times.  Our results indicate that, as the memory size grows, the quantum advantage is spread over a larger amount of time.  
	This tradeoff achieves the classical limit for spins of infinite size, for which the advantage disappears and the memory can be accessed infinitely many times.  
	
At the fundamental level, our results provides the first example of a quantum memory advantage   in a {\em deterministic} learning task involving unitary gates as the target operations.   Advantages of quantum memories have been known for longer time  for  {\em non-deterministic} learning tasks, where the learning machine has a non-zero probability of aborting. For example, Refs. \cite{nielsen1997programmable,vidal2002storing,hillery2002probabilistic,vidal2002storing,brazier2005probabilistic,ishizaka2008asymptotic,bartlett2009quantum} provide examples of machines that learn an unknown unitary gate without errors, albeit with a non-unit probability of success.     In all these examples, a quantum memory is necessary in order to achieve error-free learning.   In practice, however, no real machine is error-free, and in order to  experimentally demonstrate the advantage of the quantum memory  one needs a benchmark that quantifies the best performance achievable with classical machines.  No such benchmark has been derived  for the non-deterministic learning tasks considered in Refs. \cite{nielsen1997programmable,vidal2002storing,hillery2002probabilistic,vidal2002storing,brazier2005probabilistic,ishizaka2008asymptotic,bartlett2009quantum},  and a rigorous demonstration of the advantage of the quantum memory has not been possible so far.   A promising direction of future research is to apply the techniques developed in this paper to the derivation of quantum benchmarks for non-deterministic learning of unitary gates.

Our work calls for the experimental demonstration of quantum-enhanced learning of rotations around an unknown direction. For small values of the spin, a possible testbed is provided by NMR systems, where spin-spin interactions are naturally available \cite{vandersypen2005nmr}.   
    % In this setting, the program and the data would be different spins within the same molecule.    
    Another possibility is to use quantum dots, where one can engineer a coupling between a  single spin and an assembly of spins effectively behaving as a single spin $j$ particle  \cite{chesi2015theory}. This scenario, named the \emph{box model}, can be achieved through a uniform coupling of a central spin to the neighbouring sites.  
      No matter what platform is adopted,  our results provide the  rigorous benchmark that can be used to validate the successful demonstration of quantum-enhanced unitary gate learning in realistic scenarios where the implementation is subject to noise and experimental imperfections.

  \medskip 
  
	{\bf Acknowledgements.}  	 The authors thank E Bagan for discussions and feedback on an earlier version of the  manuscript.    This work is supported by the  National Natural Science Foundation of China through grant 11675136, the Hong Kong Research Grant Council through Grant No. 17326616 and 17300317, the Croucher Foundation,   the HKU Seed Funding for Basic Research, the Foundational Questions Institute through grant FQXi-RFP3-1325,  and the  Canadian Institute for Advanced Research (CIFAR). 
    %This project/publication  was made possible through the support of a grant  from the John Templeton Foundation. The opinions expressed in this publication are those of the author(s) and do not necessarily reflect the views of the John Templeton Foundation   

    \bibliographystyle{apsrev4-1}
    \bibliography{ProgrammingRotations}

\appendix

	\section{Derivation of the optimal quantum strategy}\label{app:lagrange}   
	In order to find the maximum of the fidelity  (\ref{fidfid}) under the constraints  (\ref{Choi-coefficient'}) we use the method of Lagrange multipliers, setting $\alpha =  x^2$ and $\beta=  y^2$.   The search of the stationary points of the fidelity yields the following four cases:

\medskip 

	\begin{enumerate}
	\item[] {\em Case 1:  $x=y=0$.}  In this case, the fidelity is given by 
	\begin{align}\label{FidOptimal}
		F_2^{\rm (e)}(j,\theta)     =       \frac1 {(2j+1)^2}  \,  {\left(    \left|   \cos  \frac\theta 2 \,   (j+1)   +  i\sin \frac \theta 2  \,  m_\theta \right|  +  \left|   \cos \frac \theta 2  \,   j    -  i  \sin \frac \theta 2  \,   m_\theta  \right|   \right)^2}   \, ,  
	\end{align}
	and is attained by  the   Choi operator 
	\begin{align}
	C^*_{\theta}  =     P_j  \otimes   |v\>\<v|  \, ,    
	\end{align}
    with 
    \begin{align}
	    |v\>   =  \sqrt{\frac{2j+2}{2j+1}}  \,e^{i\theta_+} \, |+\>  + e^{i\theta_-} \sqrt{\frac{j}{2j+1}}  \, |-\>  \, , \qquad \theta_+  =   \arctan  \left(  \frac {j}  {j+1}   \tan  \frac{\theta}2\right)  \, , \quad  \theta_-   =   -\frac{\theta}{2}  \, . 
	\end{align}
The maximum of the fidelity is attained by  $m_\theta=j$, independently of $\theta$.   Explicitly, the maximum fidelity is 
	\begin{align}\label{cuno}
		F_2^{\rm (e)}(j,\theta)     =    \frac{1 + \sqrt{ 1  +  \cos^2 \frac \theta 2   \,  \frac{2j+1}{j^2}}   +   \cos^2 \frac \theta 2   \,  \frac{2j+1}{2j^2} }{2 \, (1+ \frac 1{2j})^2}  \, .  
	\end{align} 
 		Note that the fidelity converges to 1 in the large $j$ limit, meaning that the learning becomes  nearly  perfect for large spins.   Comparison with  {\em Cases 2,3, and 4} in the following shows  that the fidelity  (\ref{cuno}) is  optimal for every angle $\theta$ whenever the spin is larger  than 1.  \\ 

	\item[] {\em Case 2:   $x\not =0,y = 0$.}   In this case, the Lagrangian  method yields the fidelity  
	\begin{align}
		F_2^{\rm (e)}(j,\theta)     = \frac{j+1}{2j+3}  \, |a|^2   \left[ 1  +  \frac{j}{j+1} \, \frac  {|c_-|^2}{  \frac{ 2j+1}{2j+3}|a|^2  -  |c_+|^2  }    \right] \, ,
	\end{align}
	achieved by setting 
	\begin{align}\label{appA:v+}
		|v_+|= \sqrt{\frac {2j}{2j+1}} \,  \frac {|c_+c_-| }{   \frac{2j+1}{2j+3}    |a|^2  -  |c_+|^2  } \, ,
	\end{align}
	and $x$ according to Eq. (\ref{Choi-coefficient'}).    The fidelity does not tend to $1$ in the large $j$ limit, indicating that the {\em Case 2}   strategy is suboptimal for large $j$.  Still, it turns out that for $j=1/2$ this strategy is optimal for some values of the angle $\theta$ around $\theta  = \pi$.  In this case, the entanglement fidelity becomes 
		\begin{align}
			F_2^{\rm (e)}\left(j=1/2,\theta\right)  =  \frac{ 1-\cos \theta}8  \,  \left [  1  - \frac 1{3(1+ 2\cos \theta)  }\right]
	\end{align}
	and the optimal Choi operator is 
	\begin{align}
	C_\theta^*     = \alpha P_{3/2}\oplus   \left(  P_{1/2}\otimes |v\>\<v|  \right) 
	\end{align}
	with 
	\begin{align}\label{47}
	\begin{array}{lll}
	 \alpha  =  \frac{1  +  8  \cos \theta + 9  (\cos \theta)^2}{3(1+2\cos \theta)^2}\, , \qquad & \qquad |v_+|   =  \sqrt{  \frac 32  \left(1-  \frac 43  \alpha\right)}\, , \qquad & \qquad \arg (v_+)  =   \arctan  \left(  \frac {1}  {3}   \tan  \frac{\theta}2\right)\\
	 &\qquad |v_-|   = \sqrt{\frac  12} \, , & \qquad \arg (v_-)  =-   \frac \theta 2 \, .
	\end{array}
	\end{align}
%For example, in the $\theta =  \pi$ case one obtains fidelity $1/3$, while the fidelity of the {\em Case 1} strategy is only $1/4$.   
	The transition from the {\em Case 1} strategy to the {\em Case 2} strategy occurs when the distance   $|\pi - \theta|$ is below the critical value  $\delta_{\rm c}  =  \arccos  [(4  +  \sqrt 7)/9]\approx 0.236\pi $.   \\

	\item[]{\em Case 3: $x\not = 0$,  $y\not =  0$.}     Note that a strategy with $y\not = 0$  can only exist for $j>1/2$, because for $j=1/2$ there  is no subspace with spin $j-1$, and therefore the coefficient $y$ is not present.  The method of Lagrange multipliers implies that, among the strategies with $x\not =  0$ and $y\not = 0$,  the maximum fidelity is attained when $x$ and $y$ take their maximum values. The corresponding the Choi operator $C_\theta^*$ is 
	\begin{align}
C_\theta^*  =  	\dfrac{2j+2}{2j+3}\,  P_{j+1}  +  \dfrac{2j}{2j-1}	\, .
\end{align}
and its fidelity is
	\begin{align}
		F_2^{\rm (e)}(j,\theta)     =     \,  \sin^2 \frac \theta 2  \, \left[  \frac{  (j+1)^2 -  m_\theta^2}{(2j+1)(2j+3)} +   \frac{  j^2 -  m_\theta^2}{(2j+1)(2j-1)}    \right] \, .           
	\end{align}
  The maximum, attained for $m_{\theta}=0$, is
	\begin{align}
		F_2^{\rm (e)}(j,\theta)     =     \,  \sin^2 \frac \theta 2  \,  \frac{ 2j^2 + 2j  -1}{(2j+3)(2j-1)} \, .
	\end{align}
	The fidelity does not reach 1 in the large $j$ limit, indicating that the {\em Case 3} strategy is suboptimal for large $j$.  Nevertheless, we find out that for $j=1$ the {\em Case 3} strategy is optimal for rotation angles around $\theta=  \pi$. % while the  {\em Case 1} strategy is optimal for the remaining values.  
		For $j=1$, the entanglement fidelity  is  
	\begin{align}\label{Fid1}
		F^{\rm (e)}_{\rm 2}(j=1,\theta)     =    \frac 35 \,  \left(\sin \frac \theta 2\right)^2   \, . 
	\end{align}
	% For   $\theta=  \pi$,  the  {\em Case 2} fidelity is $3/5$ while the fidelity for the {\em Case 1} strategy is just $4/9$.    For the other values of $\theta$, 
	A numerical comparison with the fidelity for {\em Case 1} indicates that the above fidelity is optimal for  $|\theta-\pi |   \le   \delta_{\rm c}$, with  $\delta_{\rm c}   =  0.23\pi$.        For $|\pi-\theta|  > \delta_{\rm c}$, instead, the {\em Case 1} strategy is optimal.

\iffalse	
The optimal operator $C_\theta^*$ is a linear  combination of the two projectors $P_{2}$ and $P_{0}$, corresponding to the values $2$ and $0$ of the total angular momentum---specifically, one has 
	\begin{align}
		C_\theta^*   =    \frac{ 4}{5}\,  P_{2}  + 2   \,  P_{0} \, .
	\end{align}
 \fi

	\item[]  {\em Case 4:  $x=0, y\not=0$. }   This case is similar to {\em Case 3},  and the fidelity has the expression  
	\begin{align}
	  	F_2^{\rm (e)}(j,\theta)     = \frac{j}{2j-1}  \, |b|^2   \left[ 1  +  \frac{j+1}{j} \, \frac  {|c_+|^2}{  \frac{ 2j+1}{2j-1}|b|^2  -  |c_-|^2  }    \right] \, .
	\end{align}
	By comparison with the other cases, we find that the {\em Case 4} fidelity is never optimal.  

	\end{enumerate}
\medskip  

%In summary, for $j > 1$,  the {\em Case 1} strategy is optimal for all possible rotation angles. 
% For  $j\le 1$,  the {\em Case 1} strategy is still optimal for rotation angles below a critical angle, while for larger angles the {\em Case 2} and {\em Case 3} fidelities become optimal for $j=1$ and $j=1/2$, respectively.   

Note that for Problem \ref{prob:channelonly} with $j=1$, only Case 1 and Case 2 need to be considered as the memory state is $|1,1\>$. It is easy to check that $|v_+|$ in Eq. (\ref{appA:v+}) does not  satisfy constraint Eq. (\ref{Choi-coefficient'}) for arbitrary $\theta$, showing that Case 1 is always the optimal solution for Problem \ref{prob:channelonly} when $j = 1$.

	\section{Heisenberg interaction is the optimal learning strategy}\label{app:physical}
    In this section, we prove that the channel $\map{C}_{\theta,\rm Hei}$ in Eq. (\ref{map_rel_C_theta36}) with unitary gate $U_\theta $ in  Eq. (\ref{optimalU}) is the optimal learning channel.
    To this purpose,  we calculate its entanglement fidelity $F^{(\rm e)}_{\rm Hei}(j,\theta)$, and show that it is equal to the optimal entanglement fidelity given by Eq. (\ref{FidOptimal}).

	First of all, we  note that the unitary gate $U_\theta$ can be expanded  as  
	\begin{align}
		 U_\theta  =  e^{i  h(\theta)}  \, \left  [   e^{-if(\theta)} \,   P_{j+\frac 12}  +    P_{j-\frac 12}  \right]  \, ,             
	\end{align}
	where $h  (\theta)$ is an irrelevant global phase, which we will ignore from now on.     Using this expression, we obtain the relations 
	\begin{align}
		\nonumber U_{\theta}  |j,j\>  |\h,\h\>      =   e^{-   if  (\theta)} \,   &|j,j\>  |\h,\h\>  \\
		U_{\theta}  |j,j\>  |\h,-\h\>      =  \left(  1   +  \frac  { e^{-i  f(\theta)  }-1}{2j+1}\right) \,    |j,j\>  |\h,-\h&\>  +        \frac  { \sqrt{2j} \, \left( e^{-i  f(\theta)}  - 1 \right)}{2j+1}\,       |j,j-1\>  |\h,\h\> \, ,
	\end{align}
	and we can get:
	%compute the  action of the gate $U_\theta$ on the maximally entangled state:   explicitly, we have
	\begin{align}\label{new-B3}
		\nonumber  (U_\theta \otimes I_{\rm R})  \,  (   |j,j\> \otimes     |\Phi^+\> )   =  &|j,j\> \otimes \left[   \frac{e^{-   if  (\theta)} }{\sqrt 2}   |\h,\h\> |\h,\h\>          +     \left(  1   +  \frac  { e^{-i  f(\theta)  }-1}{\sqrt 2(2j+1)  }\right) \,    |\h,-\h\>  |\h,-\h\>         \right] +  \\
	    &    +\frac  { \sqrt{j} \, \left( e^{-i  f(\theta)}  - 1 \right)}{2j+1}\,       |j,j-1\>  |\h,\h\> |\h,-\h\> \, .
	\end{align}
	
	The entanglement fidelity for this physical realization can be written as 
	\begin{align}\label{new-B4}
	F^{(\rm e)}_{\rm Hei}(j,\theta) = \<\Phi^+ |    ( V_\theta \otimes I_{\rm R})^\dag \,\left[   \left(  \map C_{\theta, {\rm Hei}}  \otimes \map I_{\rm R}\right)  \left(  |j,j\>\<j,j| \otimes   \Phi^+   \right)\right]  \,     (  V_\theta\otimes I_{\rm R}) |\Phi^+\> \, ,
	\end{align}
	where $\map I_{\rm R}$ being the identity map on the reference system ${\rm R}$.
	Then by inserting Eq. (\ref{new-B3}) and
	\begin{align}\label{new-B5}
		\left(  \map C_{\theta, {\rm Hei}}  \otimes \map I_{\rm R}\right)  \left(  |j,j\>\<j,j| \otimes   \Phi^+   \right) = \Tr_{{\rm P}_j}  \big[   (U_\theta \otimes I_{\rm R})  (|j,j\>\<j,j|  \otimes \Phi^+ )  (U_\theta^\dag \otimes I_{\rm R}) \big] \, .
	\end{align}
	into Eq. (\ref{new-B4}), we can get that:
	\begin{align}\label{new-B6}
		F^{\rm (e)}_{\rm Hei}(j,\theta)= \ \nonumber&\dfrac{1}{2(1+2j)^{2}}[1+2j+4j^{2}+2j\cos f(\theta) + (1+2j)\cos\theta+2j(1+2j)\cos(\theta-f(\theta))]\\
		\leqslant \ &\dfrac{1}{2(1+2j)^{2}}[2j^{2}+\dfrac{2j+1}{2}+\dfrac{2j+1}{2}\cos\theta + j\sqrt{1+2(2j+1)\cos\theta+(2j+1)^{2}}]\, ,
	\end{align}
	where the equality will be reached when we set $f(\theta)$ equal to Eq. (\ref{ftheta}).
	It is equal to the optimal entanglement fidelity in Eq. (\ref{FidOptimal}).

	\section{Worst-case fidelity}\label{app:worstcase}
	Here we show that learning to perform target gate $V_{\theta,g}$ by using Heisenberg interaction in Eq. (\ref{map_rel_C_theta36}, \ref{optimalU}) has an error scaling in $1/j$ in terms of the {\em worst-case fidelity} (defined by Eq. (\ref{Fwc-app})).

	The worst-case fidelity is over all learning gate $g$ and over all input target states $\psi$:
	\begin{align}\label{Fwc-app'}
	F_{w, \rm Hei}(j,\theta) =  \min_{g}  \,  \min_\psi 
	\,  F_{\rm Hei}(j,\theta, g ,  \psi) \, ,
	\end{align}
	where
	\begin{align}\label{new-D2}
	F_{\rm Hei}(j,\theta,g,\psi)  =  \<  \psi|   V_{\theta,g}^\dag  \,  \map C_{\theta, {\rm Hei}}  \big( |j,j\>\<j,j|_g    \otimes \psi  \big)   \,  V_{\theta,g}  |\psi\> \, ,
	\end{align}
	is the fidelity for the simulation of $V_g$ on the specific input state $|\psi\>$, and
	\begin{align}
	\map C_{\theta, {\rm Hei}}  \left(  |j,j\>\<j,j|_g \otimes   \psi   \right) = \Tr_{{\rm P}_j}  \big[   U_\theta  (|j,j\>\<j,j|_g  \otimes \psi )  U_\theta^\dag \big] \, ,
	\end{align}
	is calculated according to the optimal physical realization.

	Note that the trace is invariant under cyclic permutations and $V_{\theta,g} = U_g V_\theta  U_g^\dag$, we can rewrite Eq. (\ref{new-D2}) as:
	\begin{align}
	F_{\rm Hei}(j,\theta,g,\psi) = \<\psi| U_{g} V_\theta^\dagger \Big(\Tr_{{\rm P}_j}\Big[U_{\theta}(|j,j\>\<j,j| \, \otimes \,  U_{g}^\dagger|\psi\>\<\psi|U_{g})U_\theta^{\dagger}\Big]\Big) V_\theta U_{g}^\dagger |\psi\> \, .
	\end{align}

	By expanding $U_{g}^\dagger|\psi\>$ in basis \{$|\h,\h\>$, $|\h,-\h\>$\}: $U_{g}^\dagger|\psi\> = \cos\frac{\alpha}{2}|\h,\h\> + e^{i\beta}\sin\frac{\alpha}{2}|\h,-\h\>$, we find that:
	\begin{align}\label{new-D4}
	\nonumber U_{\theta}\Big(|j,j\> \otimes U_{g}^\dagger|\psi\>\Big) = &\cos\frac{\alpha}{2}|j,j\>\otimes|\h,\h\> +e^{i\beta}\sin\frac{\alpha}{2}\Big[ \frac{1}{1+2j}|j,j\>\otimes|\h,-\h\> + \dfrac{\sqrt{2j}}{1+2j}|j,j-1\>\otimes|\h,\h\> \Big]\\
	&+e^{i\beta}\sin\frac{\alpha}{2}e^{i(\theta-\frac{\sin\theta}{2j})}\Big[ \frac{2j}{1+2j}|j,j\>\otimes|\h,-\h\> -\dfrac{\sqrt{2j}}{1+2j}|j,j-1\>\otimes|\h,\h\> \Big] \, .
	\end{align}

	By inserting Eq. (\ref{new-D4}) into Eq. (\ref{new-D2}), we can get
	\begin{align}
	F_{\rm Hei}(j,\theta, g ,  \psi) = 1 -\sin^{4}\frac{\alpha}{2}\cdot\dfrac{1-\cos\theta}{j}+O\left(\dfrac{1}{j^2}\right) \, ,
	\end{align}
	showing that
	\begin{align}
	F_{w, \rm Hei}(j,\theta) = 1 -\dfrac{1-\cos\theta}{j}+O\left(\dfrac{1}{j^2}\right) \, .
	\end{align}

	\section{Proof of Theorem \ref{theo:optimalMOstrategy}}\label{app:optMOstructure}
	
The proof of the  first two items of Theorem \ref{theo:optimalMOstrategy} is identical of the proof of Lemma \ref{lem:jm}.   

It remains to prove that there exists an optimal MO strategy consisting of a covariant POVM   $(  P_{\theta, \hat g})$ and of conditional operations $\map O_{\theta,  {\hat{g}}}  =  \map U_{\hat{g}} \circ \map O_{\theta} \circ \map U_{\hat{g}}^\dag$.  

	The MO fidelity for Problem \ref{prob:optimiseall} can be expressed as 
	\begin{align}\label{FMO}
       	F_{\rm 2,MO}(j,\theta) =  \sum_y  \,  \int \d g  \,  \int \d\psi  ~\<\phi |   U_g^{(j) \dag}   P_{\theta, y}    \,  U_g^{(j)}|\phi\>  ~  \<\psi|   V_{\theta,g}^\dag  \, \map O_{\theta,  y}  (|\psi\>\<\psi|)    \, V_{\theta,g} |\psi\> \ .
    \end{align} 
    For every $y\in\set Y$, we define the probability  
    \begin{align}
	    q_{\theta, y}   =  \frac{ \Tr[P_{\theta, y}]}{2j+1}  \, ,  
    \end{align}
    the POVM  
    \begin{align}
		P_{\theta, \hat{g}}^{(y)}   :  =  \frac{  \map U_{\hat{g}}^{(j)\dag}  (P_{\theta,y})}{  q_{\theta, y}}   \, , 
	\end{align}
	and the quantum channels  
	\begin{align}\label{Cthetaxg}
		\map O_{\theta, y,  {\hat{g}}}  :  =   \map U^\dag_{\hat{g}}  \circ \map O_{\theta, y} \circ \map U_{\hat{g}}   \, .   
    \end{align}    
	Note that  the operators  $\left( P_{\theta, \hat{g}}^{(y)}\right)_{g\in\set G}$  satisfy the normalization condition 
	\begin{align}
		\int \d \hat{g}  \,   P_{\theta, \hat{g}}^{(y)}   =  I_{{\rm P}_j}   \qquad \forall y  \in \set Y \, ,
	\end{align}    
	following from Schur's lemma.  

	In terms of the above probabilities, POVMs, and channels, the expression (\ref{FMO})  can be rewritten as 
	\begin{align}
	    \nonumber   	F_{\rm 2,MO}(j,\theta)   &=  \sum_y   q_{\theta, y}    \,  \int \d g  \,  \int \d\psi  ~\<\phi |  \,  P^{(y)}_{\theta, g}   \, |\phi\>  ~  \<\psi| \,   U_g  V_\theta^\dagger  U_g^\dag    \,   \map O_{\theta,y}       (|\psi\>\<\psi|)    \,   U_g V_\theta  U_g^\dag |\psi\>  \\
		\nonumber 	  &=  \sum_y   q_{\theta, y}    \,  \int \d g  \,  \int \d\psi'  ~\<\phi |  \,  P^{(y)}_{\theta, g}   \, |\phi\>  ~  \<\psi'|   V_\theta^\dagger   \,  U_g^\dag   \map O_{\theta,y}       (U_g  |\psi'\>\<\psi'| U_g^\dag)      U_g \,  V_\theta |\psi'\>  \\
	    &=  \sum_y   q_{\theta, y}   \,  \int \d g  \,  \int \d\psi'  ~\<\phi |  \,  P^{(y)}_{\theta, g}   \, |\phi\>  ~  \<\psi'|   V_\theta^\dagger   \,    \map O_{\theta,y, g}       (  |\psi'\>\<\psi'|)     \,  V_\theta |\psi'\>  	 \ .
    \end{align} 
	Since the fidelity is a convex combination,  we have the upper bound 
	\begin{align}\label{rhs}
       	F_{\rm 2,MO}(j,\theta) \le     \max_y    \Big\{    \int \d g  \,  \int \d\psi  ~\<\phi |  \,  P^{(y)}_{\theta, g}   \, |\phi\>  ~  \<\psi| V_\theta^\dagger \,    \map O_{\theta,y, g}    (|\psi\>\<\psi|)  V_\theta  \, |\psi\> \Big\}  \, . 
    \end{align}   
	It is immediate to check that the bound is    attained by the MO strategy  consisting of the POVM  $\left(P_{\theta, g}^{(y_*)}\right)_{g\in \grp G}$ and of the conditional operations $
    \map O_{\theta, y_*,g}$, where $y_*$ is the outcome that maximizes the expression in the right-hand-side of Equation (\ref{rhs}).

	\section{Optimization of the MO strategy}\label{app:MOoptimization}
	Our goal is to maximize the fidelity 
    \begin{align}
	    \nonumber F^{\rm (e)}_{\rm 2,MO}(j,\theta) &= (2j+1)  \,  \int  \d g  \,     | \<  \xi_\theta      |j,m\>_{g} |^2 ~    \big|  \<   \Phi_{V_{\theta^{'}}}^+   |\Phi^+_{\theta,g}\> \big|^2\\
	    \nonumber &= (2j+1) \, \int \d g \, | \< \xi_\theta | U_g^{(j)} |j,m \> |^2 ~    \big|  \<   \Phi_{V_{\theta^{'}}}^+ | (V_{\theta,g} \otimes I)  |\Phi^+\> \big|^2\\
	    \nonumber &= (2j+1) \, \int \d g \, | \< \xi_\theta | U_g^{(j)} |j,m \> |^2 ~    \big|  \<   \Phi_{V_{\theta^{'}}}^+ | (U_g V_\theta U_g^\dagger \otimes I) |\Phi^+\> \big|^2\\
	    &= (2j+1) \, \int \d g \, | \< \xi_\theta | U_g^{(j)} |j,m \> |^2 ~    \big|  \<   \Phi_{V_{\theta^{'}}}^+ | (U_g \otimes \overline{U}_g) |\Phi^+_{V_\theta}\> \big|^2 \, ,
	    %&=   (2j+1)  \,  \int  \d g  \,     | \<  \xi_\theta    |  U_g|  j,m\> |^2 ~    \big|  \< \Phi^+_\tau  |    (U_g\otimes \overline U_g)|\Phi_{\theta}^+    \> \big|^2    \, , 
	\end{align}  
    over all values of $m$, all unit vectors $\xi_\theta$, and all unitary gates $V_{\theta^{'}}$.
    Using the relation  $\overline U_g =  \sigma_y  U_g\sigma_y$, we can rewrite the fidelity as 
    \begin{align}
	    F^{\rm (e)}_{\rm 2,MO}(j,\theta) &= (2j+1) \, \int \d g \, | \< \xi_\theta | U_g |j,m \> |^2 ~    \big|  \<   \Phi_{V_{\theta^{'}}}^* | (U_g \otimes U_g) |\Phi^*_{V_\theta}\> \big|^2    \, , 
	\end{align}  
    with  $|\Phi_{V_{\theta^{'}}}^*\>  =   (I\otimes \sigma_y) \,  |\Phi_{V_{\theta^{'}}}^+\>$ and $|\Phi^*_{V_\theta}\>  =   (I\otimes \sigma_y) \,  |\Phi^+_{V_\theta}\>$.

	For every angle $\alpha $, the vector    $|\Phi^*_{V_\alpha}\>$  can be expanded as 
	\begin{align}
	\nonumber  |\Phi^*_{V_\alpha}\>    & =     i   \cos\frac \alpha 2  \,  \left|0,0; \frac 12 ,\frac 12\right\>  + \sin\frac \alpha 2  \, \left |1,0 ;\frac 12 ,\frac 12 \right\>   \, ,
	\end{align}
	having used the notation $  |l,n;  j_1,  j_2\>$ for the eigenstates of the $z$-component of the total spin of a bipartite system consisting of two spins $j_1$ and $j_2$, respectively. 
	   Hence,  we have 
	\begin{align}
		\nonumber \< \Phi_{V_{\theta^{'}}}^* | (U_g \otimes U_g) |\Phi^*_{V_\theta}\>   &=     \cos  \frac{\theta^{'}}{2}  \cos \frac \theta 2    +      \sin \frac{\theta^{'}}{2}      \sin \frac\theta 2   \,    \<1,0|      U^{(1)}_g  |1,0\>    
	\end{align}
	and 
	\begin{align}
		\nonumber \Big | \< \Phi_{V_{\theta^{'}}}^* | (U_g \otimes U_g) |\Phi^*_{V_\theta}\> \Big|^2     =   &     \cos^2 \frac{\theta^{'}}{2}      \cos^2 \frac \theta 2    +  \frac 13  \sin^2 \frac{\theta^{'}}{2}      \sin^2  \frac\theta 2   \\
		\nonumber 	& +  2   \cos \frac{\theta^{'}}{2}      \cos \frac \theta 2      \sin \frac{\theta^{'}}{2}      \sin \frac\theta 2   \,    \<1,0|      U^{(1)}_g  |1,0\>   \\
		\label{fourier} & + \frac 23 \, \sin^2 \frac{\theta^{'}}{2}      \sin^2  \frac\theta 2      \,     \< 2,0|   U_g^{(2)}  | 2,0\>  \, .
	\end{align}
	Moreover, the fidelity can be expressed as  
	\begin{align}
		\nonumber      F^{\rm (e)}_{\rm 2,MO}(j,\theta) &  =  (2j+1)\,  \int \d g \,    \Big | \< \Phi_{V_{\theta^{'}}}^* | (U_g \otimes U_g) |\Phi^*_{V_\theta}\> \Big|^2   ~  \Big|\<\xi_\theta  |  U_g^{(j)}|j,m\>\Big|^2\\
	   \nonumber   &  =  (2j+1)\,    \int \d g \,    \Big | \< \Phi_{V_{\theta^{'}}}^* | (U_g \otimes U_g) |\Phi^*_{V_\theta}\> \Big|^2   ~  \< \xi_\theta  |  U_g^{(j)}|j,m\>  ~  \<\overline \xi_\theta  |  \overline U_g^{(j)}|j,m\>    \\
	   \nonumber   &  =  (2j+1)\,    \int \d g \,    \Big | \< \Phi_{V_{\theta^{'}}}^* | (U_g \otimes U_g) |\Phi^*_{V_\theta}\> \Big|^2   ~  \<  \xi_\theta  |  U_g^{(j)}|j,m\>  ~  \<\overline \xi_\theta  |  e^{-i \pi  J_y}\,  U_g^{(j)}    e^{i \pi   J_y} |j,m\>    \\
	   &  = (-1)^{j-m} \, (2j+1)\,    \int \d g \,    \Big | \< \Phi_{V_{\theta^{'}}}^* | (U_g \otimes U_g) |\Phi^*_{V_\theta}\> \Big|^2   ~ \big (   \<  \xi_\theta  |\otimes  \<\widetilde \xi_\theta|  \big) \,  \left( U_g^{(j)} \otimes  U_g^{(j)} \right)  \,  \big  (  |j,m\> \otimes |j,-m\> \big) \, ,
	\end{align}
	having defined	 $|\widetilde \xi_\theta\>      =    e^{i \pi    J_y } \,  |\overline \xi_\theta\>$\, .  We now insert Equation (\ref{fourier}) into the above expression,  taking advantage of the orthogonality relation 
	\begin{align}
	\int  \d g    \,   \overline{ \<  l,n|     U^{(l)}_g  |l,n'\>}~  \left ( U_g^{(j_1) } \otimes U_g^{(j_2) }  \right)    =  \frac 1{2l+1}  \,  |l,n; j_1,j_2\>\<l,n';  j_1,j_2| \, .
	\end{align}
	In this way, the fidelity becomes 
	\begin{align}
		\nonumber      F^{\rm (e)}_{\rm 2,MO}(j,\theta) &  =  (-1)^{j-m} \,    (2j+1)\,    (   \<  \xi_\theta  |\otimes  \<\widetilde \xi_\theta|  \big) \,\Gamma  \,  \big  (  |j,m\> \otimes |j,-m\> \big) \, ,
	\end{align}	
	with 
	\begin{align}\label{gamma}
		\nonumber \Gamma    =  &   \Big( \cos^2 \frac{\theta^{'}}{2}      \cos^2 \frac \theta 2   +  \frac 13  \sin^2 \frac{\theta^{'}}{2}      \sin^2  \frac\theta 2\Big) \,    |0,0; j,j\>\<0,0;  j,j|  \\
		\nonumber    &+   \Big(  \frac 23   \cos \frac{\theta^{'}}{2}      \cos \frac \theta 2      \sin \frac{\theta^{'}}{2}      \sin \frac\theta 2 \Big)  \,  |1,0; j,j\>\<1,0;  j,j|  \\
		 &+   \Big( \frac 2{15} \, \sin^2 \frac{\theta^{'}}{2}      \sin^2  \frac\theta 2 \Big) \,   |2,0; j,j\>\<2,0;  j,j| 
		   \, .  
	\end{align}
	Expanding $|\xi_\theta\>$ as $|\xi_\theta  \>   =  \sum_{n}  \, \xi_{\theta, n}  \, |j,n\>$, we obtain 
	\begin{align}
		\nonumber F^{\rm (e)}_{\rm 2,MO}(j,\theta) &  =   (2j+1)\,   \sum_n   \,  |\xi_{\theta, n}|^2   \,  (-1)^{n-m} ~   (   \<  j,n  |\otimes  \<j,-n|  \big) \,\Gamma  \,  \big  (  |j,m\> \otimes |j,-m\> \big)\\
	& \le (2j+1)\,\max_n  \,    (-1)^{n-m} ~   \big(   \<  j,n  |\otimes  \<j,-n|  \big) \,\Gamma  \,  \big  (  |j,m\> \otimes |j,-m\> \big) \, .
	%	\\		  &   \le   (2j+1)\,\max_{m,n}   \Big|   \,   \big(\<  j,n  |\otimes  \<j,-n|  \big) \,\Gamma  \,  \big  (  |j,m\> \otimes |j,-m\> \big)    \Big|   \, .   
		    \label{lkj}
	\end{align}
	Note that the bound can be attained by choosing $|\xi_\theta\>$ to be an eigenstate of $J_z$ with suitable eigenvalue $n$. 
		  
	Now, let $|\Gamma|  = \sqrt{  \Gamma^2}$  be the the modulus of $\Gamma$, and let $ \Gamma_+   :=  ( |\Gamma|+\Gamma)/2 $ and $ \Gamma_-   :=  (|\Gamma|-  \Gamma)/2 $ be the positive and negative part of $\Gamma$, respectively. 
	With inserting these definitions in Eq. (\ref{lkj}), the fidelity can be upper bounded as
	\begin{align}
		\nonumber F^{\rm (e)}_{\rm 2,MO}(j,\theta)		\nonumber &    \le   (2j+1)\,\max_{n}   \Big|   \,   \big(\<  j,n  |\otimes  \<j,-n|  \big) \,  \left( \sqrt  {\Gamma_+}   +  \sqrt  {\Gamma_-} \right)   \,  \left( \sqrt  {\Gamma_+}   -  \sqrt  {\Gamma_-} \right) \,  \big  (  |j,m\> \otimes |j,-m\> \big)    \Big|     \\
		\nonumber & \le  (2j+1)\,\max_{n}  \sqrt{   \big(\<  j,n  |\otimes  \<j,-n|  \big) \,|\Gamma | \,  \big  (  |j,n\> \otimes |j,-n\> \big) ~  \big(\<  j,m  |\otimes  \<j,-m|  \big) \,|\Gamma | \,  \big  (  |j,m\> \otimes |j,-m\> \big) }  \\
		\label{Gamma} &  \le  (2j+1)\,\max_m        \big(\<  j,m  |\otimes  \<j,-m|  \big) \,|\Gamma | \,  \big  (  |j,m\> \otimes |j,-m\> \big)   \, ,
	\end{align}
	the second inequality following from the Cauchy-Schwarz inequality applied to the vectors  $\left( \sqrt  {\Gamma_+}   -  \sqrt  {\Gamma_-} \right) \,  \big  (  |j,m\> \otimes |j,-m\> \big) $ and $\left( \sqrt  {\Gamma_+}   +  \sqrt  {\Gamma_-} \right) \,  \big  (  |j,n\> \otimes |j,-n\> \big)$.   We will discuss the attainability of the bound (\ref{Gamma}) in the end of the proof.

	Inserting  the  definition of $\Gamma$   [Eq. (\ref{gamma})]  in the bound (\ref{Gamma}), we obtain 
	\begin{align}
		\nonumber \big(\<  j,m  |\otimes  \<j,-m|  \big) \,|\Gamma|  \,  \big  (  |j,m\> \otimes |j,-m\> \big)   =   &    \Big( \cos^2 \frac{\theta^{'}}{2}      \cos^2 \frac \theta 2   +  \frac 13  \sin^2 \frac{\theta^{'}}{2}      \sin^2  \frac\theta 2\Big) \,    \left|  \<0,0;  j,j| \,  \big  (  |  j,m \> \otimes  |j,-m\>\right)  \, \big|^2 \\
		\nonumber    &+   \Big|  \frac 23   \cos \frac{\theta^{'}}{2}      \cos \frac \theta 2      \sin \frac{\theta^{'}}{2}      \sin \frac\theta 2 \Big|  \,      \left|  \<1,0;  j,j| \,  \big  (  |  j,m \> \otimes  |j,-m\>\right)  \, \big|^2  \\
		 &+   \Big( \frac 2{15} \, \sin^2 \frac{\theta^{'}}{2}      \sin^2  \frac\theta 2 \Big) \,   
		     \left|  \<2,0;  j,j| \,  \big  (  |  j,m \> \otimes  |j,-m\>\right)  \, \big|^2 \, , 
	\end{align}
	which becomes 
	\begin{align}
		\nonumber 	  \nonumber \big(\<  j,m  |\otimes  \<j,-m|  \big) \,|\Gamma | \,  \big  (  |j,m\> \otimes |j,-m\> \big)         = &  	    \Big( \cos^2 \frac{\theta^{'}}{2}      \cos^2 \frac \theta 2   +  \frac 13  \sin^2 \frac{\theta^{'}}{2}      \sin^2  \frac\theta 2\Big) \,  \frac 1 {2j+1}
		 \\
		 \nonumber    &+   \Big|  \frac 23   \cos \frac{\theta^{'}}{2}      \cos \frac \theta 2      \sin \frac{\theta^{'}}{2}      \sin \frac\theta 2 \Big|  \,\frac{3 m^2}{j (j+1)  (2j+1)}\\
		 &+   \Big( \frac 2{15} \, \sin^2 \frac{\theta^{'}}{2}      \sin^2  \frac\theta 2 \Big) \,   
		    \frac{ 5 \big(  j^2  + j  - 3 m^2\big)^2}{j (j+1)  (2j-1) (2j+1)  (2j+3)}   \, .  \label{gammadilegno}
	\end{align}
	
	For $j>1$, one can easily see that each of the three summands in the above expression has its maximum value for $|m |=   j$, independently of the angles $\theta$ and $\theta'$. 
	Setting $m=j$ and optimizing over $\theta'$ we obtain that the maximum is obtained for 
	\begin{align}\label{tauinapp2}
		\theta^{'} =\arccot \left[ \cot \theta  +  \dfrac{2j+1+ 2\cos\theta}{(2j^{2}+3j)\sin\theta} \right] \, ,
	\end{align}
	for $\theta$ in $[0,\pi]$, and by 
	\begin{align}\label{tauinapp3}
		\theta^{'} =\pi+  \arccot \left[ \cot \theta  +  \dfrac{2j+1+ 2\cos\theta}{(2j^{2}+3j)\sin\theta} \right] \, ,
	\end{align}
	for $\theta$ in  $(\pi,2\pi)$ (recall that the range of $\arccot$ is   between $0$ and $\pi$).  For these values of $\theta^{'}$,  the entanglement fidelity  is
    \begin{align}
		F_{\rm  2,MO, opt}^{(\rm e)}&(j,\theta)   = \dfrac{(2j+1)(1+\cos(\theta-\theta^{'}))}{2(2j+3)} +\dfrac{(2j+1)(\cos\theta+\cos\theta^{'})+\cos(\theta+\theta^{'})+1}{2(j+1)(2j+3)}  \, .
    \end{align}
    The same approach works for $j=1/2$, in which case $|m|  = j $ is the only possible choice, and the optimization over $\theta'$ yields again the optimal value  (\ref{tauinapp2}). 
    
	Note that the choice of angles $\theta'$    in Eqs. (\ref{tauinapp2}) and (\ref{tauinapp3}) satisfies the condition $ \cos \frac{\theta^{'}}{2}      \cos \frac \theta 2      \sin \frac{\theta^{'}}{2}      \sin \frac\theta 2  \ge 0$.  
   Hence, the operator $\Gamma$ is positive, and therefore $\Gamma  = |\Gamma|$.   As a consequence, the inequality  (\ref{Gamma}) is attained by choosing $|\xi_\theta\>   =    |j,m\>$.

    For $j = 1$, the optimal MO strategy is  determined by a brute-force approach, by setting $m=0$ and $m=1$, optimizing the right-hand-side of Eq.  (\ref{gammadilegno}) over $\theta'$. 
       When $|\pi - \theta| > 0.303\pi$, the optimal MO strategy is the same as when $j \neq 1$.
    When $|\pi - \theta| \leqslant 0.303\pi$, the optimal $m$ is $m=0$, and the optimal angle $\theta^{'}$ becomes  $\theta^{'} = \pi$. Also in this case, the operator $\Gamma$ is positive, and therefore the inequality  (\ref{Gamma}) is attained by choosing $|\xi_\theta\>   =    |j,m\>$.

    \iffalse
	In terms of  average input-output fidelity, we obtain  the value
    \begin{align}
	    \nonumber   &  F_{\rm  MO}(j,\theta)_{\rm opt}    = \dfrac{4j+4+(2j+1)\cos(\theta-\tau)}{6j+9} \\
		& +\dfrac{(2j+1)(\cos\theta+\cos\tau)+\cos(\theta+\tau)+1}{3(j+1)(2j+3)}  \, .
    \end{align}
    The maximum fidelity is achieved by using the program state $U_{g(\textbf{n})}|j,j\>$, measuring   the coherent state POVM $\{P_{{\textbf n'}}=U_{g({\textbf n'})}|j,j\>\<j,j|U_{g({\bf n'})}^{\dagger}\}$, and rotating around the axis $\bf n'$ of the angle $\tau$ determined by  Eq. (\ref{tauinapp2}).   Note that the angle $\tau$ converges to $\theta$ in the large $j$ limit. 

\fi

	\section{Persistence of the quantum advantage}\label{app:longevity}
	The state of the memory spin after the interaction can be obtained by application of the complementary channel $ \widetilde{\map C}_\theta $, defined by 
    \begin{align}\label{recycle}
        \widetilde{\map C}_\theta   (\rho^{(j)})   =    \Tr_{\rm S} \left[     U_\theta   \,   \left ( \rho^{(j)} \otimes  \frac I 2  \right)   U^\dag_{\theta} \right]  \, ,
    \end{align}
    where $\Tr_{{\rm S}}$ denotes the partial trace over the target spin, and $U_\theta$ is the unitary operator in Eq. (\ref{optimalU}).
    
    To evaluate this state, it is convenient to look at the evolution of the basis states $ |j,m\>_{g}:  = U_g^{(j)}  |j,m\>$.  By explicit calculation, we obtain the relation  
    \begin{align}
        \widetilde{\map C}_\theta  \Big( |j,m\>\<j,m|_{g}   \Big)   =  \sum_{i=-1}^{1}  c_{m+i,m}    \,   |j,m+i\>\<j,m+i |_{g}   \ ,
    \end{align}
    where the coefficients $c_{m+i,m}$ are given by 
    \begin{align}
       	\begin{cases}
       		c_{m-1,m}=&\nonumber\dfrac{(j+m)(1+j-m)}{(1+2j)^{2}}(1-\cos\theta-\dfrac{\sin^{2}\theta}{2j})\\
       		c_{m,m}=&\nonumber 1-c_{m-1,m}-c_{m+1,m}\\
       		c_{m+1,m}=&\nonumber\dfrac{(j-m)(1+j+m)}{(1+2j)^{2}}(1-\cos\theta-\dfrac{\sin^{2}\theta}{2j})
       	\end{cases} \ ,
    \end{align}

    At the first step, the memory starts in the state $|j,j\>_{g}$.  By repeatedly applying Eq.(\ref{recycle}), we then obtain the memory state at every step.  Explicitly, the memory state for the $n$-th usage  is given by  
    \begin{align}\label{staten-1}
        \widetilde {\map C}_\theta^{n-1}    \Big(  |j,j\>\<j,j|_{g}\Big)    =   \sum_{m=   j-n+1}^j \,    p(n-1,m,\theta)  \,   |j,m\>\<j,m|_{g}  \, , 
    \end{align}  
    where $p(n-1,m,\theta)$ is the probability distribution after $n-1$ usages, which is given by 
    \begin{align}\label{recycling-distribution}
       	p(n,m,\theta)=&\nonumber \sum_{i=j-m}^{n}(-1)^{i+j-m}
       	\begin{pmatrix}
       		n \\ i
       	\end{pmatrix}
        \begin{pmatrix}
       		i \\ j-m
       	\end{pmatrix}\frac{i!}{(\frac{2j}{1-\cos\theta})^{i}}\\
       	=&(-1)^{j-m+1}\,  \frac{2j}{(1-\cos\theta)}  \,  \frac{n!}{(n-j+m)!}  \times U\left(j-m+1,n+2,-\frac{2j}{1-\cos\theta} \right) \, ,
    \end{align}
    $U$ being  Tricomi's function (confluent hypergeometric function of the second kind). 
Using the recursion formula
 \\
	\begin{align}
	    U(a,b,z) = (2a-b+z+2)U(a+1,b,z) -(a+1)(a-b+2)U(a+2,b,z) \, ,
	\end{align}
	we  get the asymptotic expression 
	\begin{align}\label{equ-D8}
		p(n,m,\theta)= \dfrac{2j}{n(1-\cos\theta)+2j}\cdot   \left[\dfrac{n(1-\cos\theta)}{n(1-\cos\theta)+2j}\right]^{j-m} + O\left(\frac{1}{j}\right) \, .
	\end{align}

 Now,    Equation (\ref{staten-1}) gives us the memory state  at the $n$-th iteration.    The fidelity obtained by using this state is given by 
    \begin{align}\label{(equ-D1)}
       	F_{\rm Hei}(j,\theta,n) =\sum_m  \,  p(n-1,m,\theta)    ~  F_{\rm Hei}(j,\theta,m) \ ,
    \end{align}
    where $F_{\rm Hei}(j,\theta,m)$ is the average fidelity when the probe is in the state $|j,m\>_g$, namely 
    \begin{align}\label{FHei}
	    F_{\rm Hei} (j,\theta,  m)=  \int \d g \int \d \psi ~ \<\psi|V_{\theta,g}^{\dag}  \, \Tr_{{\rm P}_j} \Big[   U_\theta \big( |j,m\>_{g} \<j,m|_{g}    \otimes  \psi   \big) U_\theta^\dag \Big]  \,  V_{\theta,g}|\psi\> \, ,
    \end{align}  
 The average over the input states can be easily computed using the relation with the entanglement fidelity, Equation (\ref{horodecki}) . 
Using Equation  (\ref{map_rel_C_theta36}) for the gate $U_\theta$,  we obtain the asymptotic expression
    \begin{align}\label{equ-D4}
       	F_{\rm Hei}(j,\theta,m)=1- \dfrac{(1+2j-2m)(1-\cos\theta)}{3j}+O\left(\frac{1}{j^{2}}\right) \ .
    \end{align}

	One can see directly that in asymptotics, $F(j,\theta,m)$ is a arithmetic progression and $p(n,m,\theta)$ is a geometric progression.
	Inserting the above  expressions into Eq. (\ref{(equ-D1)})   we obtain  
	\begin{align}\label{recyclingfidelity}
		F_{\rm Hei}(j,\theta,n) = 1 - \dfrac{1-\cos\theta}{3j}\cdot\dfrac{n(1-\cos\theta)+j}{j} + O\left(\frac{1}{j^{2}}\right)   \, .
	\end{align}
	Comparing with the MO fidelity in Eq.(\ref{FMOapp}), we obtain that the persistence  of the quantum advantage  tends to $N(j,\theta)=j/(1-\cos\theta)$.
	
	\begin{figure}[ht]
		\centering
		\includegraphics[width=0.51\textwidth]{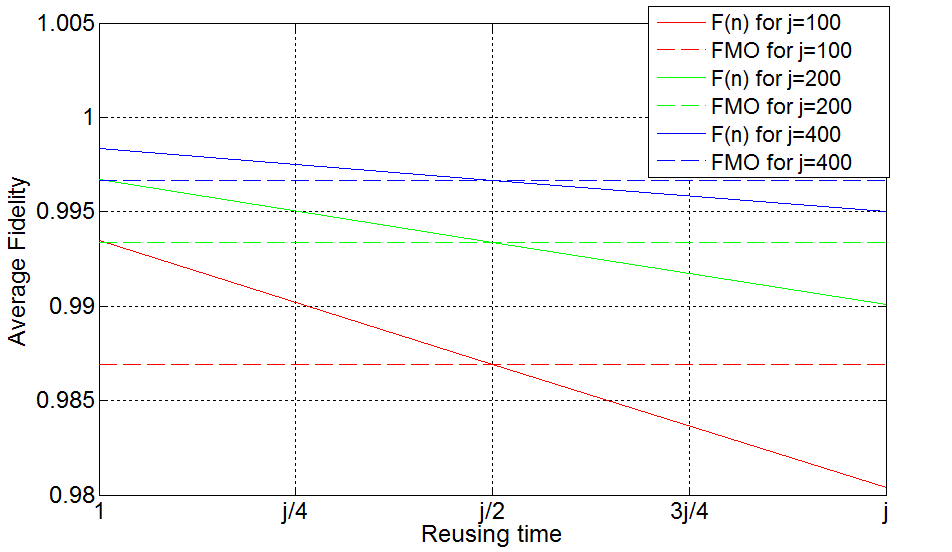}
		\caption{\footnotesize
			\textbf{Degradation of the fidelity with the number of recycling steps.}   The dependence of the fidelity on the number  $n$ of recycling steps is plotted for $j=100$  (red solid line),   $j=200$  (green solid line), and $j =400$ (blue solid line), in the case of rotation angle $\theta=  \pi$.    The plot shows an inverse linear scaling  with the recycling step $n$.     The dotted lines give the values of the MO fidelities for $j=100$  (red), $j=200$  (green), and $j=400$ (blue).   The  fidelity of this protocol falls under the MO fidelity when the number of recycling steps is larger than $j/2$. }
		\label{fig5}
	\end{figure}
	The exact dependence of  the fidelity on $n$ is shown  in Figure \ref{fig5} for different values of the spin and for rotation angle $\theta= \pi$. 
	Interestingly, the persistence of the quantum advantage is exactly equal to the asymptotic value $j/2$ for all the values of $j$ shown  in the figure.  
	     
	We showed the explicit  calculation of  $F(j,\theta,m)$ and $p(n-1,m,\theta)$ when the interaction time is fixed at every step. 
	More general strategies where  the interaction time is  optimized at every step can be studied in the same way.  In the large $j$ limit, we find that such step-by-step optimization is not needed:  the  fidelity  tends to the same value, no matter whether the interaction time is optimized at every step or once for all. As a result, the persistence of the quantum advantage is the same in both scenarios.

	\section{Robustness of the quantum strategy}\label{app:robustness}
Here we evaluate the fidelity in the execution of the gate    $V_{\theta,\st n}  =  \cos \frac \theta 2   \,  I  - i\sin \frac \theta 2  \,  \st n\cdot \bs \sigma$ when the optimal learning strategy for pure states is adopted with a probe in the thermal state $\rho_{\st n, \gamma}$. 
The fidelity of this strategy is 
\begin{align}
	    F_{\rm Hei} (j,\theta,  \gamma):=  \int \d \st n \int \d \psi ~ \<\psi|V_{\theta,\bf n}^{\dag}  \, \Tr_{{\rm P}_j} \Big[   U_\theta \big( \rho_{\st n,\gamma}    \otimes  \psi   \big) U_\theta^\dag \Big]  \,  V_{\theta,\bf n}|\psi\> \, ,
    \end{align}  
with $U_\theta$ as in Equation  (\ref{optimalU}).  Inserting the expression for the state $ \rho_{\st n, \gamma}$ into the above equation, we obtain 
		\begin{align}
		F_{\rm Hei}(j,\theta,\gamma)=  \frac {\sinh \gamma}{\sinh[ (2j+1) \gamma]} \, \sum_m  \,  e^{2\gamma m}     ~  F_{\rm Hei}(j,\theta,m) \, , 
			\end{align}
			with 	    $F_{\rm Hei} (j,\theta,  m)$ defined as in Equation (\ref{FHei}).  The asymptotic expression for $F_{\rm Hei}(j,\theta,m)$ was computed  in  Equation (\ref{equ-D4}).    Inserting this expression in the above equation,  we obtain  
\begin{align}
	\nonumber 	F_{\rm Hei}(j,\theta,\gamma)  &=  \frac {\sinh \gamma}{\sinh[ (2j+1) \gamma]} \, \sum_m  \,  e^{2\gamma m}     ~ \left [ 1- \dfrac{(1+2j-2m)(1-\cos\theta)}{3j}+O\left(\frac{1}{j^{2}}\right) \right] \\
	 &  	  
	=  1 - \dfrac{1-\cos\theta}{3j \tanh \gamma} + O\left(\dfrac{1}{j^2}\right) 	   \, . 
\end{align}
  
    \section{Learning higher dimensional rotations for spin-$k$ particle}\label{app:spink}
    Following the structure of the optimal learning mechanism for spin $1/2$,  we choose the memory state to be $|j,j\>_{g}$ and we let the two spins undergo  the Heisenberg interaction
    \begin{align}
        U_\theta^{(k)}  =     \exp  \left[    -i \theta \, \frac{  \,   2\st J \cdot \st K}{2j+1}  \right] \, ,
    \end{align} 
    where $\st K  =  (K_x, K_y,  K_z)$ are the spin operators of the target spin.
    
	Using the above strategy,  we can explicitly compute the  entanglement fidelity, given by 
    \begin{align}\label{D2}
        F_{\rm Hei}^{(\rm e)}(j,k,\theta) = \<\Phi^{(k)+} |    ( V_\theta^{(k)} \otimes I_{\rm R})^\dag \,\left[   \left(  \map C^{(k)}_{\theta, \rm Hei}  \otimes \map I_{\rm R}\right)  \left(  |j,j\>\<j,j| \otimes   \Phi^{(k)+}   \right)\right]  \,     (  V_\theta^{(k)}\otimes I_{\rm R}) |\Phi^{(k)+}\> \, ,
    \end{align}
    \begin{align}
	    \left(  \map C_{\theta, \rm Hei}^{(k)}  \otimes \map I_{\rm R}\right)  \left(  |j,j\>\<j,j| \otimes   \Phi^{(k)+}   \right) = \Tr_{{\rm P}_j}  \big[   (U_\theta^{(k)} \otimes I_{\rm R})  (|j,j\>\<j,j|  \otimes \Phi^{(k)+} )  (U_\theta^{(k)\dag} \otimes I_{\rm R}) \big] \, ,
    \end{align}
    where $|\Phi^{(k)+}\> = \frac{1}{2k+1}\sum_{m=-k}^k |k,m\>\otimes |k,m\>$ being the canonical maximally entangled state of two spin-$k$ particles, $\rm R$ denotes a reference qubit, entangled with the target spin-$k$ particle, and $V_\theta^{(k)}$  is a rotation of $\theta$ around the $z$ axis in $2k+1$ representation.
    
	Inserting the formula of $U_{\theta}^{(k)}$ in Eq. (\ref{D2}), using the expressions of the Clebsch-Gordan coefficients, 
    we  arrive at the asymptotic expression 
    \begin{align}
	    F_{\rm Hei}^{\rm (e)}(j,k,\theta)= 1-\dfrac{2k(k+1)}{3j}(1-\cos\theta) \ + \ O\left(\frac{1}{j^{2}}\right) \, .
    \end{align}
	The average fidelity is then given by
    \begin{align}\label{app:G5}
	    F_{\rm Hei}(j,k,\theta)= 1-\dfrac{k(2k+1)}{3j}(1-\cos\theta) \ + \ O\left(\frac{1}{j^{2}}\right) \, .
    \end{align}
        
	A similar calculation can be done for the  MO strategy consisting in measuring the memory state with POVM $P_{\hat g}  =  (2j+1)\,   \map U_{\hat g}^\dag  (|j,j\>\<j,j|)$ and then performing the conditional operation  $V_{\theta,\hat g}^{(k)} = U_{\hat g}^{(k)}V_\theta^{(k)}U_{\hat g}^{(k)\dagger} $ on the target spin-$k$ particle, which means rotate with angle $\theta$ with the rotated $z$-axis $\hat{g}\bf{e}_z$:     
	\begin{align}
	    F^{\rm (e)}_{\rm MO}(j,k,\theta)\nonumber
	    =\int d{g}\int d{\hat g}\Tr\left[P_{\hat g}\phi_{g}\right]F^{\rm (e)}(j,k, \theta, g, {\hat g}) 
    \end{align}
    with
    \begin{align}
	    F^{\rm (e)}(j,k,\theta, g, {\hat g}) =
	    \dfrac{1}{(2k+1)^{2}}\left|\Tr\left[V_{\theta,{g}}^{{(k)}\dagger}V_{\theta,{\hat g}}^{(k)}\right]\right|^{2} \, .
    \end{align}
    
    By denoting $\varphi$ as the angle between $|j,j\>_g$ and $|j,j\>_{\hat g}$, and  $\tau$ the rotation angle for the rotation   $V_{\theta,{g}}^{(k)\dagger}V_{\theta,{\hat g}}^{(k)}$, the entanglement fidelity can be rewritten as 
    \begin{align}
	    F^{\rm (e)}_{\rm MO}(j,k,\theta)=\dfrac{\int_{0}^{\pi}d\varphi \sin\varphi (\cos\varphi)^{4j}\dfrac{\sin^{2}(\frac{2k+1}{2}\tau)}{\sin^{2}\frac{\tau}{2}}}{(2k+1)^{2}\int_{0}^{\pi}d\varphi \sin\varphi (\cos\varphi)^{4j}} \, . 
    \end{align}
	Performing the average, we obtain the asymptotic expression        
	\begin{align}
	    F^{\rm (e)}_{\rm MO}(j,k,\theta) = 1-\dfrac{4k(k+1)}{3j}(1-\cos\theta) +O\left(\frac{1}{j^{2}}\right) \, ,
    \end{align}
    which can then be used to evaluate the average fidelity as
    \begin{align}
	    F_{\rm MO}(j,k,\theta) = 1-\dfrac{2k(2k+1)}{3j}(1-\cos\theta) +O\left(\frac{1}{j^{2}}\right) \ .
    \end{align}
    By comparing with Eq. (\ref{app:G5}), we again see that the error is exactly twice the error of the coherent quantum learning strategy.

\end{document}